\documentclass{article}

\usepackage{arxiv}

\usepackage{tikz}
\usepackage{tabularx}
\usepackage[utf8]{inputenc}
\usepackage{subcaption}
\usepackage[frozencache]{minted}
\usepackage{algorithm}
\usepackage{algpseudocode}
\usepackage{caption}
\usepackage{csquotes}
\usepackage{hyperref}
\usepackage{comment}
\usepackage{diagbox}
\usepackage{graphicx}
\usepackage{wrapfig}
\usepackage{float}
\usepackage{url}
\usepackage[british]{babel}
\usepackage{bm}
\usepackage{amsmath}
\usepackage{amssymb}
\usepackage{amsthm}
\usepackage{textcomp}
\usepackage{gensymb}

\begin{document}

\title{Call graph discovery in binary programs from unknown instruction set architectures}

\author{Håvard Pettersen \\
	Department of Computer Science\\
	Norwegian University of Science and Technology\\
	Trondheim, Norway\\
\texttt{haavapet@stud.ntnu.no} \\
\And%
Donn Morrison \\
	Department of Computer Science\\
	Norwegian University of Science and Technology\\
	Trondheim, Norway\\
	\texttt{donn.morrison@ntnu.no}}

\maketitle              % typeset the header of the contribution
\begin{abstract}
    This study addresses the challenge of reverse engineering binaries from unknown instruction set architectures, a complex task with potential implications for software maintenance and cyber-security. We focus on the tasks of detecting candidate call and return opcodes for automatic extraction of call graphs in order to simplify the reverse engineering process. Empirical testing on a small dataset of binary files from different architectures demonstrates that the approach can accurately detect specific opcodes under conditions of noisy data. The method lays the groundwork for a valuable tool for reverse engineering where the reverse engineer has minimal a priori knowledge of the underlying instruction set architecture.

\keywords{Reverse engineering \and unknown CPU architecture \and program call graph \and binary analysis}
\end{abstract}
\section{Introduction}
In an era defined by rapid technological advancements and a vast amount of different systems, further amplified by the rise of the Internet of Things, the importance of understanding and decoding the inner workings of software cannot be understated. At the heart of this is the process of reverse engineering. Reverse engineering in the context of software, is the practice of inspecting, deconstructing, and analysing the structure and operation of a binary file in order to understand its architecture, design, and functionality. This is often done without access to source code or design documentation, making it a painstaking, yet critical, part of software analysis and security.

The reverse engineering process is notably important in areas such as cyber-security, where detecting and understanding malware is key to developing and maintaining robust security. It also plays a vital role in maintaining and debugging legacy software and firmware, where the original documentation or developers may not be available. For these reasons, reverse engineering is a critical skill in the digital age and an important area in need of further research and development efforts.

In the broader context, several tools and methods have been developed over time to aid the reverse engineering process. Most of these tools require a priori knowledge about the instruction set architectures of the binary being analysed, which poses limitations and challenges when dealing with unknown or undocumented instruction set architectures.

The current methods for reverse engineering binaries from unknown instruction set architectures are limited and often involve invasive procedures such as hardware decapsulation, which can be costly, slow, and potentially damaging to the hardware \cite{hardware}. Additionally, obfuscation measures are often used to deliberately make the process even more challenging and time-consuming. Examples of such techniques are custom virtual machines used to execute the binary file \cite{xu2018vmhunt,kinder2012towards}.

When looking at the process of reverse engineering from a methodological perspective, a common practice is detecting important functions and focusing the reverse engineering efforts on them, so-called sub-routine scanning \cite{observational}. Hence, a tool capable of generating call graphs for binaries would alleviate much of the needed efforts in the current reverse engineering process.

There is a clear need for heuristic tools that can assist reverse engineers in extracting meaningful information from such binaries without prior knowledge of the instruction set architectures. With this in mind, the following research questions are formulated:
\begin{enumerate}
    \item [RQ1] Can a call graph be heuristically deduced from binary programs of an unknown instruction set architecture? \label{RQ1}
    \item [RQ2] How effective is the heuristic approach and what are its limitations? \label{RQ2}
\end{enumerate}

The central contribution of this study is the development and validation of a method to detect opcodes and generate call graphs from binaries with unknown instruction set architectures. Our method is evaluated in detail, revealing its capabilities and limitations. A secondary contribution is a metric called the Opcode Candidacy Probability Score (OCP-Score). This metric enables the ranking of opcodes based on likelihood, showing the reverse engineer the most probable call-return pairs.
    
The structure of the rest of the paper is as follows: Section \ref{background} describes the background and related work. Section \ref{methodology} describes the methodology and proposed algorithm. Section \ref{results} evaluates the proposed approach on a small dataset of binary programs from different instruction set architectures. Section \ref{discussion} offers a discussion of the results and Section \ref{conclusion} concludes and proposes potential avenues for further research.

\section{Background and related work}
\label{background}
In this section we briefly introduce the background and most relevant related work.

\subsection{Background}

An instruction set architecture serves as an abstract model of the computer on which software runs, and when compiling a program, one must target a specific instruction set architecture. This instruction set architecture defines the supported instructions, data types, addressing modes, and other relevant aspects of the architecture. Consequently, a program compiled for eg. the x86\_64 architecture will not execute on a computer with ARM architecture without the use of emulators.

Assembly code is a mnemonic of machine code, meaning there is a one-to-one mapping between them. For instance, an instruction \ttfamily mov r1 \#2 \normalfont could be assembled into the following bytes: \ttfamily 0x5e83a2\normalfont. In much the same way, disassembly would mean translating the bytes back to the original assembly instructions. Typically, an instruction consists of an opcode, which specifies the operation, and operands, which determine the values to operate on. These operand values can include memory addresses, immediate values, or registers.

The instruction format demarcates the bits of an instruction representing the opcode, and the bits representing the operands. An instruction format can either be fixed length, where all instructions are the same length, or variable length (e.g., x86\_64 architecture). Table \ref{tab:mips_instruction_format} illustrates the fixed-width instruction format for the MIPS architecture. Additionally, the endianness of the instruction set architecture is an important consideration, indicating the order in which bytes are stored. An instruction stored as \ttfamily 0x1234 \normalfont would be represented as \ttfamily 0x1234 \normalfont for big-endian, and \ttfamily 0x3412 \normalfont for little-endian.

\begin{table}
\caption[Instruction format of the MIPS architecture]{Instruction format of the MIPS architecture, illustrating the arithmetic (R-type), immediate (I-type), and jump (J-type) instruction formats \cite{mipsinstrfig}.}
    \label{tab:mips_instruction_format}
    \centering
    \begin{tabular}{c|c|c|c|c|c|c}
        \hline
        \hline
        R-type & op & rs & rt & rd & shamt & funct \\
        \hline
        I-type & op & rs & rt & \multicolumn{3}{c}{address/immediate} \\
        \hline
        J-type & op & \multicolumn{5}{c}{target address} \\
        \hline
        \hline
        Field size & 6 bits & 5 bits & 5 bits & 5 bits & 5 bits & 6 bits \\
        \hline
    \end{tabular}
\end{table}

\begin{figure}
    \centering
    \includegraphics[width=0.5\linewidth]{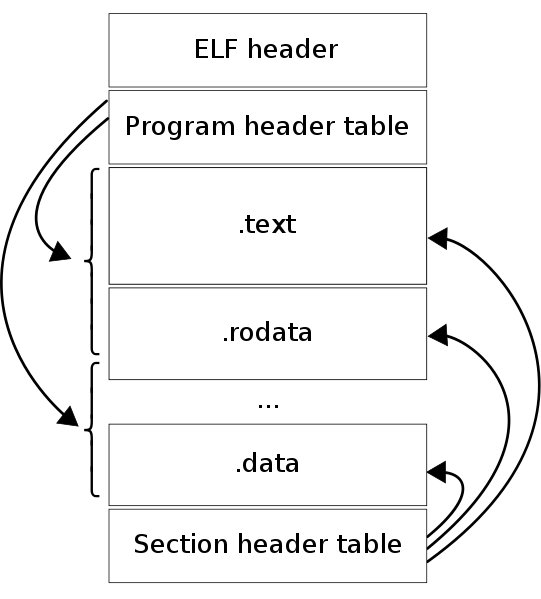}
    \caption[ELF file structure]{ELF file structure \cite{wikielf}.}
    \label{fig:elf}
\end{figure}

Compiled programs and firmware are typically stored in a binary file format, for example the Executable and Linkable Format (ELF). This is of interest in our study because a binary file often contains more than just instructions; it also contains data and metadata. In the case of ELF, there are sections and segments of different types of data. In Figure \ref{fig:elf}, which shows the contents of an ELF file, we are specifically interested in the \textit{.text} segment, as that is where the instructions are stored. When dealing with an unfamiliar file format, it is of interest to identify the start and end of the corresponding \textit{.text} segment, to accurately isolate and extract the instructions.

\begin{figure}

%\tikzset{EdgeStyle/.append style = {->}}
%\begin{tikzpicture}
%  \Vertex{main}
%  \SO(main){func1}
%  \EA(func1){func2}
%  \Edge(main)(func1)
%  \Edge(main)(func2)
%\end{tikzpicture}
    \begin{minipage}{.5\textwidth}
    \centering
    \includegraphics[width=0.5\linewidth]{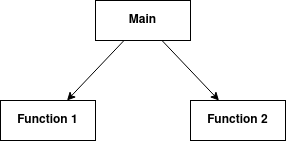}
    \end{minipage}
    \begin{minipage}{.5\textwidth}
    \begin{minted}{c}
    int main()
    {
        Function1();
        ...
        Function2();
    }
    void Function1() { ... }
    void Function2() { ... }
    \end{minted}
    \end{minipage}
    \caption[Example call graph]{Call graph constructed from a program containing a main function which calls Function 1 and Function 2.}
    \label{fig:call_graph}
\end{figure}

To detect call graphs in a binary with an unknown instruction set architecture, the most important task is detecting the function boundaries, namely the byte position at which a function starts and ends. It is typical that compiled programs from known architectures exhibit distinct function epilogues and prologues, in the form of return instructions and stack operations, respectively. An example of a call graph for a simple program, consisting of a main function that calls two other functions, can be seen in Figure \ref{fig:call_graph}. A more complex call graph may have characteristics such as cycles, which can be the result of compiled recursion and loops.

Call instructions generally reference other functions in one of three ways: absolute addressing, where the operand of the instruction is the address we want to access; relative addressing, where the operand of the instruction contains the offset from the current address; and register addressing, where the address of the callee is stored and accessed through a register. In general, it is simpler to detect where a function points when it uses absolute and relative addressing, with register addressing being difficult without runtime knowledge.

%Developing a static analysis method capable of disassembling any binary file for any architecture is an exceedingly difficult problem. However, it is feasible to create a tool that aids in the disassembly process for a subset of architectures or architectures with specific properties. In interpreted languages like Python, it is possible to create an executable using tools like PyInstaller\footnote{https://pyinstaller.org/en/stable/}. However, without prior knowledge of it being a Python executable, conducting static analysis on such files can prove challenging. This is due to both the interpreter being bundled with the executable, and the code itself being interpreted byte code. As a general rule of thumb, a program written in a lower-level language like C will be easier to extract meaningful information from than a program written in a higher-level language.

\subsection{Related work}
Reverse engineering is an active research field where applications range from malware detection \cite{bao2014application} to vulnerability detection \cite{sharif2009automatic} to architecture classification \cite{clemens2015automatic}. However, nearly all software reverse engineering research assumes a priori knowledge of the instruction set architecture upon which it is executed. Software reverse engineering without knowledge of the instruction set architecture depends first on reverse engineering the hardware, and such research is scarce.

%Many tools exist that facilitate reverse engineering of binaries from known instruction set architectures. These include IDA Pro \footnote{\url{https://www.hex-rays.com/ida-pro/}}, a widely used disassembly tool, allows interactive disassembly of binaries across popular architectures. Similarly, the Python library angr \footnote{\url{https://angr.io/}}, assists in symbolic analysis of binary files, provided the architecture of the binary is known and supported. Ghidra \footnote{\url{https://ghidra-sre.org/}} is an open-source reverse engineering suite rapidly gaining popularity and supporting a wide variety of architectures and executable formats. radare2 \footnote{\url{https://rada.re/n/radare2.html}} is a free/libre console-based toolchain supporting many reverse engineering and debugging tasks. ILspy \footnote{https://github.com/icsharpcode/ILSpy} and JD Project \footnote{\url{http://java-decompiler.github.io/}} are tools that can be used to disassemble .NET and Java binaries respectively.

Clemens \cite{clemens2015automatic} uses a dataset of 16,000 binaries from 20 different architectures to detect endianness and instruction set architecture. The approach relied heavily on byte frequency distributions as features, suggesting that they retained sufficient opcode information for accurate instruction set architecture classification. The approach is similar to the approach of Kairajarvi \textit{et al.} \cite{kairajarvi2020isadetect}, and relies on the instruction set architecture being part of the training data. 

Qiu et al. \cite{qiu2015identifying} introduce a function representation called the reverse extended control flow graph (RECFG) for function identification that doesn't rely on function prologues and epilogues. They address four key challenges in this approach: 1) difficulty in differentiating data from code when file formats are unknown; 2) sensitivity to disassembly starting points for variable-length instructions; 3) the risk of inaccurate disassembly due to candidate return instructions being part of another instruction; and 4) issues with relying on compiler-specific prologues and epilogues. Their method uses a multilayer perceptron trained with features based on the 32 bytes around a candidate opcode and debug symbols as groundtruth. However, the method requires detailed knowledge of the ISA (the candidate opcodes must be decoded from their instructions).

%Qiu et al \cite{qiu2015identifying} propose a function representation called reverse extended control flow graph (RECFG) which is used as a basis for function identification that is not dependent on function prologue and epilogues. The authors cite four main motivations: 1) firmware or executable file formats are not always known, making differentiation between data and code difficult; 2) disassembly of variable length instructions such as x86/x64 are sensitive to disassembly starting location (e.g., entry point); 3) byte(s) of a candidate return instruction could in reality be part of another instruction leading to inaccurate disassembly; and 4) dependence on well-known prologues and epilogues as function boundaries (often compiler-specific) does not handle the case of hand-crafted assembly, and would lead to poor disassembly in this special case. A multilayer perceptron ins trained using features consisting of the 32 bytes surrounding a candidate opcode and using debug symbols as a groundtruth. While the method improves upon existing techniques, it requires detailed knowledge of the ISA, and thus is not applicable to domains where this information (instruction format, opcode definitions, etc.) is unavailable.

Sharif \textit{et al.} \cite{sharif2009automatic} developed a system called Rotalume to reverse engineer binaries that have been obfuscated using programs such as VMprotect \footnote{https://vmpsoft.com/}. This approach was however dependent on executing the binary in a protected environment, in order to extract runtime information, which makes the approach unfeasible for binaries with an unknown architecture.

In an unpublished work by Chernov \textit{et al.} \cite{2012chernov}, a heuristic approach is presented, where they detect instruction set architectural features in binaries with unknown instruction set architecture. They present multiple assumptions of the binary file of an unknown architecture: Call opcodes usually have the absolute address of a function as an operand, a function prologue is closely spatially located to the previous functions epilogue, and call and return opcodes are amongst the most commonly used opcodes. Through the use of frequency distributions and address matching, they were able to detect subroutines and control flow in binaries, through only static analysis of the binary file. The work done in this report is based on the same assumptions made by Chernov \textit{et al.} but differs in its implementation. It will also be the first \textit{published} research on this specific topic.

Most studies discussed have necessitated prior knowledge of the instruction set architecture, with only the last paper presented by Chernov \textit{et al.} focusing on unknown architectures. As such there is a clear research gap identified in this area, which this paper aims to contribute towards.

\section{Methodology}
\label{methodology}
This section introduces our heuristic approach, as well as dataset acquisition and analysis strategy.

\subsection{Call graph extraction}

\begin{figure}
    \centering
    \includegraphics[width=0.9\linewidth]{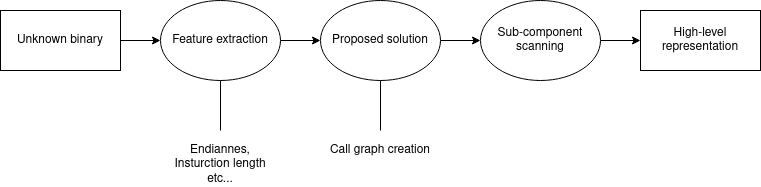}
    \caption[Context of the proposed solution]{Context of the use of the proposed solution, occurring between architectural feature extraction and sub-component scanning.}
    \label{fig:context}
\end{figure}

At a high-level, our approach\footnote{https://github.com/haavapet/binary-analysis} takes a binary file and a set of parameters as input, and returns a list of potential call graphs along ranked by probability. Figure \ref{fig:context} depicts a context where a reverse engineer would use the method as part of the process of reverse engineering a binary towards a high-level representation.

A small binary from the Chip8 architecture was used for verification during development of the method due to its instruction format being well-suited for static analysis. This enabled testing under the assumption that properties of the Chip8 ISA were not known, and then checking results against the ISA as a groundtruth. In Sections \ref{results} and \ref{discussion} we will further analyse and discuss the method against a more common and comprehensive set of architectures.

A high-level pseudo-code of the main algorithm can be seen in Algorithm \ref{algo}. The \textit{extract\textunderscore instruction} function separates the bytes of the binary into a list of instructions, based on the provided instruction length, and file offsets. The \textit{get\_potential\_edges} function finds all instructions with the given call opcode where its operand points to a valid instruction, with either relative or absolute addressing. The \textit{filter\_valid\_edges} function validates edges by confirming that the given return opcode is one of the few instructions above the called instruction, to ensure there is a distinct function epilogue followed by a function prologue.

\begin{algorithm}[H]
\caption{Detect call graph from binary}\label{algo}
\begin{algorithmic}[H]
\State instructions = extract\_instructions(...) \Comment{bytes $\rightarrow$ List[Instructions]}
\State top\_candidates = Heap(...)

\For{call candidates}
    \State potential\_edges = get\_potential\_edges(...)
    \For{return candidates}
        \State valid\_edges = filter\_valid\_edges(...)
        \State probability = get\_probability(...)
        \State store\_candidate\_in\_heap(...)
    \EndFor
\EndFor

\For{candidate in top\_candidates}
    \State create\_graph\_for\_candidate(...)
\EndFor

\State \Return candidates\_with\_graph
\end{algorithmic}
\end{algorithm}

Support for relative addressing was implemented as additional functionality and can be seen in Algorithm \ref{algo2}.

\begin{algorithm}[H]
\caption{Get potential edges - relative adressing}\label{algo2}
\begin{algorithmic}[H]
\State potential\_call\_instructions = get\_instructions\_with\_opcode(...)
\For{potential\_call\_instructions}
    \State signed\_operand = int\_to\_signed\_int(instruction.operand)
    \If{signed\_operand hits relative instruction address}
        \State add\_edge(...)      
    \EndIf
\EndFor

\State \Return edges
\end{algorithmic}
\end{algorithm}

We introduce a metric for computing a probability for a given call opcode and return opcode pair to rank candidates for visual inspection by a reverse engineer. The metric is listed as Equation \ref{prob_eq}, and is referred to as \textbf{Opcode Candidacy Probability Score} (\textbf{OCP-Score}).

\begin{equation}
\label{prob_eq}
    \textbf{OCP-Score}_{op} = \frac{a \cdot (\textbf{valid edges}_{op}) + (\textbf{potential call edges}_{op})}{b \cdot (\textbf{call count}_{op})},
\end{equation}

where $a=2$, $b=3$, and \textbf{call count}$_{op}$ refers to the number of instructions with the given call opcode ${op}$. \textbf{potential edges} refers to the number of edges associated with the candidate call opcode ${op}$, in particular, those instances where the operand points to a valid address. \textbf{valid edges} refers to the number of edges where there is a function epilogue, specifically a candidate return instruction, located within a specified range before the address of the candidate call instruction containing the opcode ${op}$.

The OCP-Score is normalised within the range $[0,1]$, explained by the constraint that \textbf{length valid edges} and \textbf{potential call edges} are always less than or equal to \textbf{call count} (numerator and denominator constants $a$ and $b$). It is worth noting that \textbf{valid edges} is weighted more heavily than \textbf{potential call edges}, due to being more strongly correlated with only call instructions as opposed to call and branch instructions. The OCP-Score will be evaluated and discussed further in Sections \ref{results} and \ref{discussion}.

To reduce the large opcode search space, the analysis of the binary file requires a set of initial parameters provided alongside the file itself. The parameters, their type, and a description can be seen in Table \ref{tab:parameters}. All parameters are currently required by the API, however, a potential modification with reasonable defaults and increased search space, could require only the first three parameters (instructionLength, retOpcodeLength, and callOpcodeLength) while keeping the rest optional, which would greatly increase usability. The reverse engineer would normally play a role in determining initial values for parameters.

{\renewcommand{\arraystretch}{1.3}%
\begin{table}
    \centering
    \captionsetup{justification=centering}
    \caption[Explanation of the API parameters]{Explanation of the API parameters.}
    \begin{tabular}{p{4cm}|c|p{6cm}}
        \textbf{Parameter} & \textbf{Type} & \textbf{Description} \\\hline\hline
        instructionLength & int & Length of an instruction in bits \\\hline
        retOpcodeLength & int & Length of instruction return opcode in bits \\\hline
        callOpcodeLength & int & Length of instruction call opcode in bits \\\hline
        fileOffset & int & Byte position of code section start in binary \\\hline
        fileOffsetEnd & int & Byte position of code section end in binary \\\hline
        pcOffset & int & Address of first instruction \\\hline
        pcIncPerInstr & int & Distance between the address of each instruction \\\hline
        endiannes & string & "big" or "little" \\\hline
        nrCandidates & int & How many graph candidates to return \\\hline
        callCandidateRange & int, int & Only search the [x:y] most popular instruction with a bitmask of callOpcodeLength as a potential call candidate \\\hline
        retCandidateRange & int, int & Only search the [x:y] most popular instruction with a bitmask of retOpcodeLength as a potential return candidate \\\hline
        returnToFunction-PrologueDistance & int & Distance from function epilogue (return instruction) to function prologue (call operand address) \\\hline
        unknownCodeEntry & bool & Search the binary for the most optimal fileOffset and fileOffsetEnd, drastically increases runtime \\\hline
        includeInstructions & bool & Include instructions in the result object. Recommended False for big binaries if rendering graph \\\hline
        isRelativeAddressing & bool & Relative or absolute addressing for call operands \\
         
    \end{tabular}
    \label{tab:parameters}
\end{table}}

\begin{figure}
    \centering
    \begin{subfigure}[b]{0.47\textwidth}
        \centering
        \includegraphics[width=\textwidth]{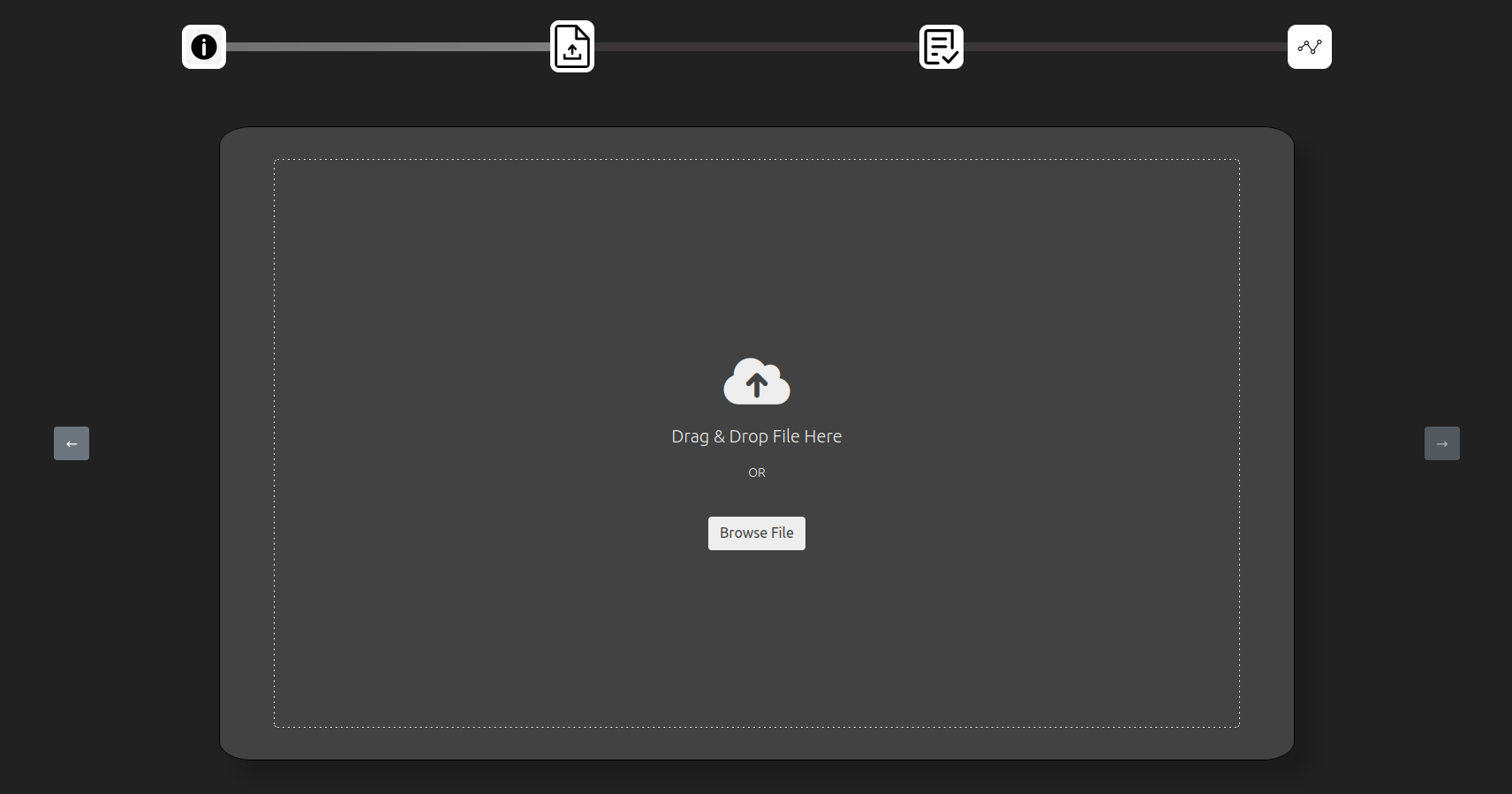}
        \caption{{\small Upload file page}}   
    \end{subfigure}
    \hfill
    \begin{subfigure}[b]{0.47\textwidth}  
        \centering 
        \includegraphics[width=\textwidth]{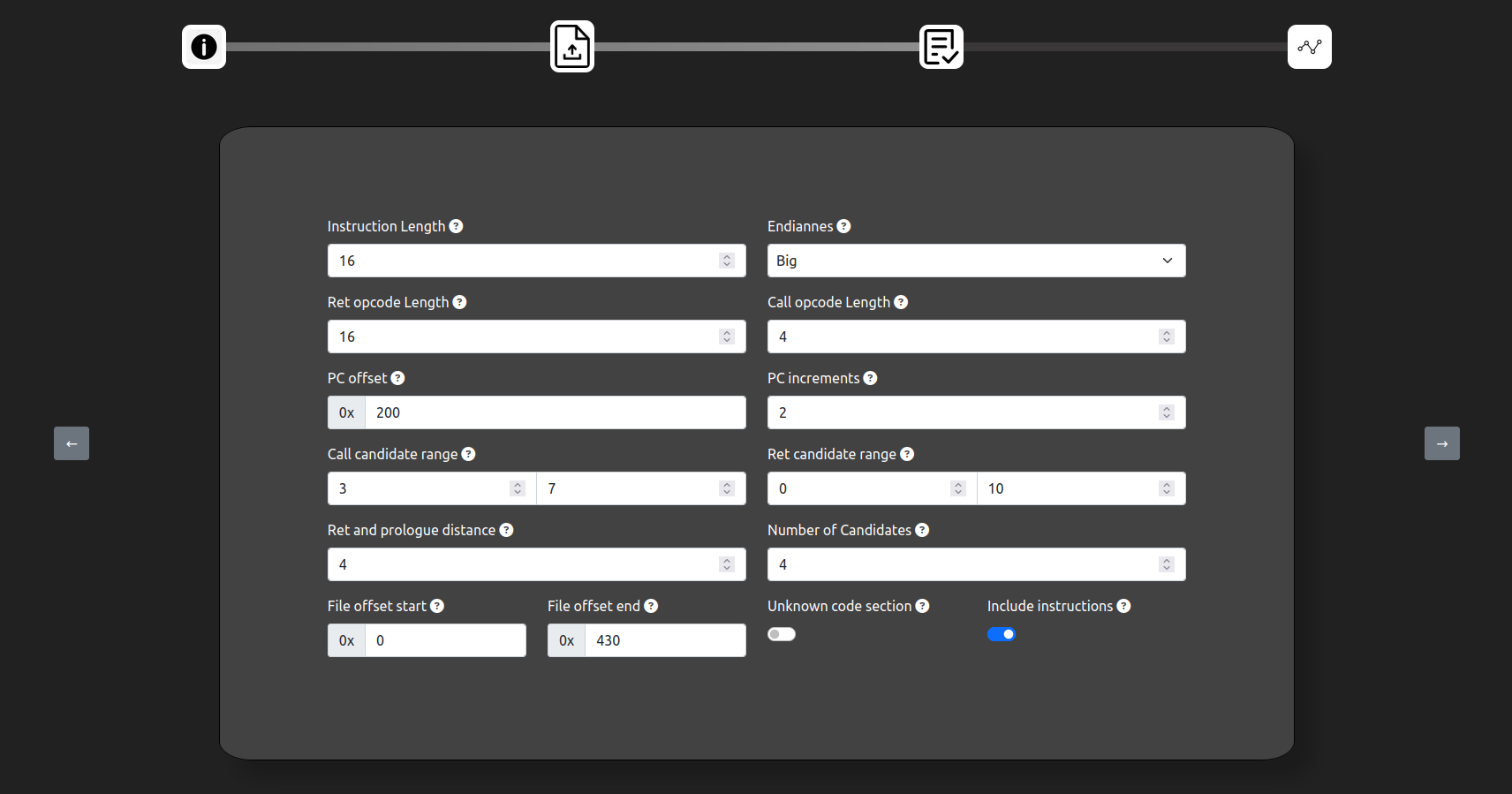}
        \caption[]%
        {{\small Form page}}    
    \end{subfigure}
    \vskip\baselineskip
    \begin{subfigure}[b]{0.47\textwidth}   
        \centering 
        \includegraphics[width=\textwidth]{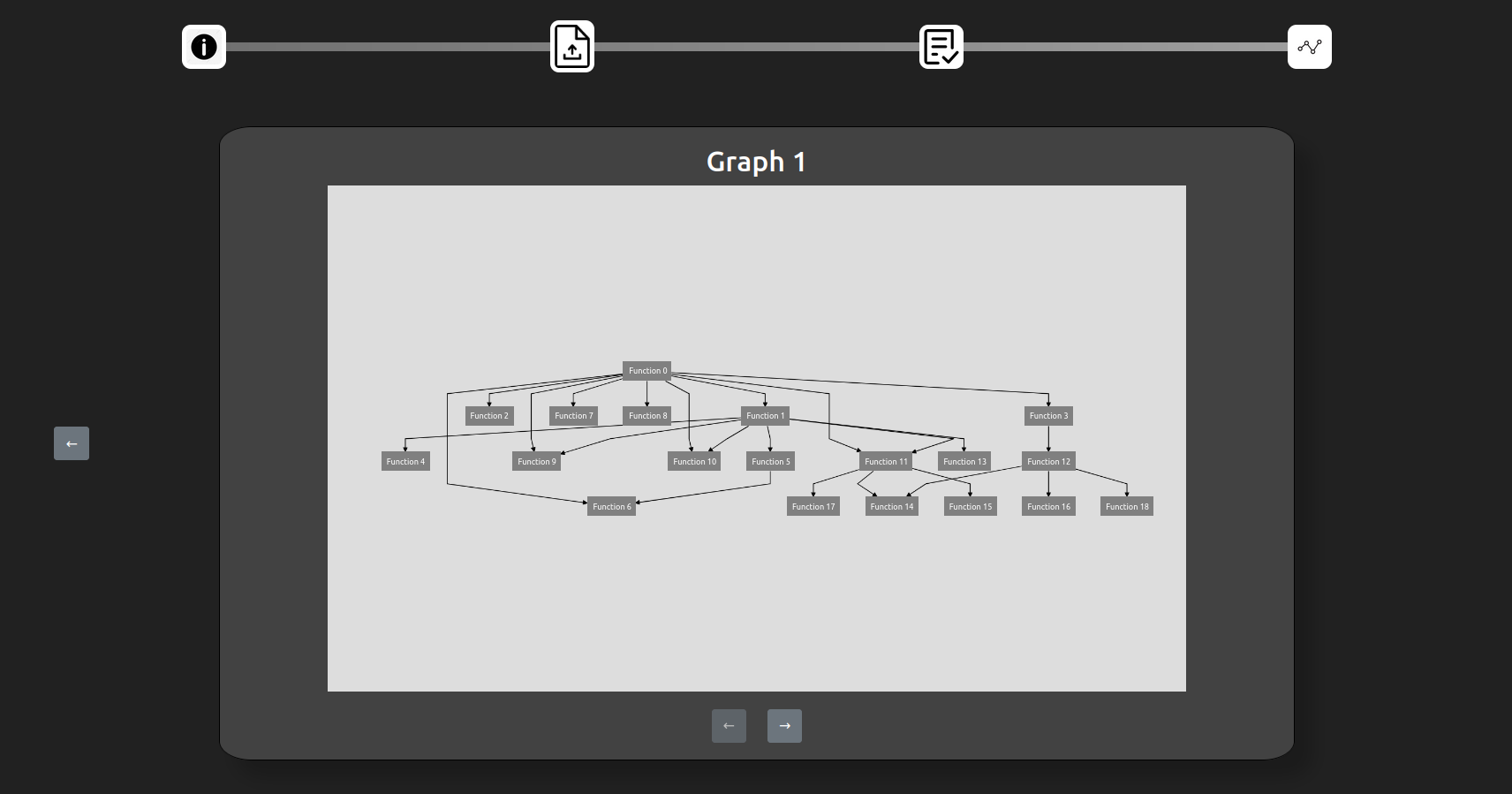}
        \caption[]%
        {{\small Display graph page}}    
    \end{subfigure}
    \hfill
    \begin{subfigure}[b]{0.47\textwidth}   
        \centering 
        \includegraphics[width=\textwidth]{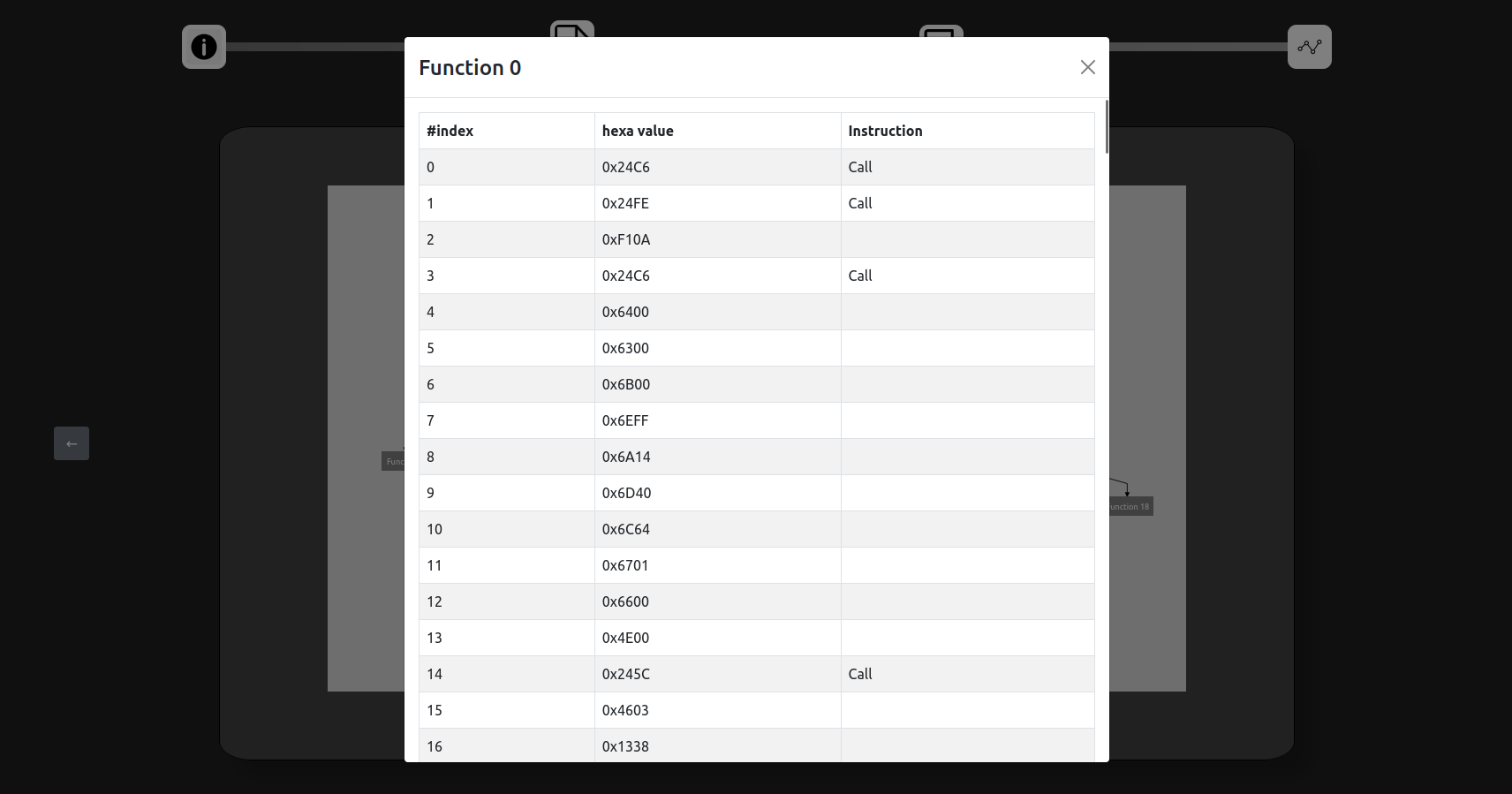}
        \caption[]%
        {{\small Modal after clicking function 0}}    
    \end{subfigure}
    \caption[Frontend user interface]{User interface of the frontend solution, showing the different pages for uploading a binary file, entering parameters, and displaying the generated call graph.} 
    \label{fig:frontend}
\end{figure}

Alongside the backend implementation in Python, a frontend written in React was also developed for ease of use. The frontend provides a simple graphical interface where the reverse engineer can upload the binary file, input the required parameters, and then display the created call graphs. Figure \ref{fig:frontend} shows the interface of the frontend.

%The development of the program has incorporated multiple practices from the XP framework, such as test-driven development, coding standards, and continuous integration \cite[p.~49]{xp}. Among the tests are end-to-end tests to validate the program as a whole, along with several tests for the different modular parts of the program, such as instruction extraction. Continuous integration through GitHub actions ensures that incremental updates to the program are validated, and pre-commit hooks ensure that faulty code is not pushed to the GitHub repository. In addition to tests, linters for both the program and the frontend, as well as static type checking for the program, are also included. This ensures that certain code standards are held, and enforces consistent code styles throughout the code. 

%The project can be run locally using either docker-compose \footnote{https://www.docker.com/}, or the programming environments' respective package managers: poetry \footnote{https://python-poetry.org/} for Python, and npm \footnote{https://www.npmjs.com/} for React. For detailed instructions, see the included README file.

\subsection{Analysis strategy and data generation}
The results and analysis focus on three integral parts of the presented heuristic method. The first part is to input the binary program with the correct parameters and ensure that the returned call opcode and return opcode are correct. The second part evaluates the assigned OCP-Score under different inputs, to detect how noisy and potentially faulty data affects the output. The third part will be looking at the created call graph of a small binary file, and comparing it to a call graph created by inspecting the source code.

There were multiple considerations taken into account when choosing programs and architectures for the opcode detection and OCP-Score evaluation analysis. Firstly, the architecture should conform to a fixed-length instruction format, as that is what our method expects and should be evaluated against. However, a reference binary with a variable-length instruction format has been included to provide insights into the behavior of the method under such conditions. Secondly, the binary should contain sufficient immediate or relative call and return instructions. Lastly, the programs used should be commonly used, complex, and written in a low-level language like C.

The most important characteristic of the binary used for the call graph creation was that the program is sufficiently small, this is to ensure the creation of a human-readable call graph, as well as reducing the manual labor required to create a call graph from inspecting the source code. In addition to this, it is important that the binary conforms to the same properties as mentioned in the previous paragraph.

%The output of the program was fed to a Python script, which uses Matplotlib \footnote{https://matplotlib.org/} to produce the graphs seen in the forthcoming section. The specific binaries used in this analysis, along with their associated parameters, are explained in detail in Section \ref{expsetup}.

\section{Results}
\label{results}
This section will analyse three important parts of the proposed solution: opcode detection, call graph creation, and the OCP-Score. In addition to this, the experimental setup will be described such that the results can be reproduced.

\subsection{Experimental setup}
\label{expsetup}
In order to reproduce the results in the following analysis, one can use the binaries in Table \ref{tab:binaries_source}, with the corresponding list of parameters found in Table \ref{tab:parameters_binaries}. 

There are seven binaries in total, and they are all included in the accompanying GitHub repository. The binaries span three different programs: cURL, OpenVPN, and Chipquarium.

Four architectures are used in the analysis. The MIPS and Aarch64 architectures conform to a fixed-length instruction format, while the x86\_64 architecture uses a variable-length instruction format. The Chipquarium binary, used in the call graph analysis, is compiled for the Chip8 architecture and is also the binary used during the development and testing of the method. 

During the analysis it was found that the cURL MIPS binary had almost no occurrence of immediate call instructions, hence a new version of cURL MIPS was cross-compiled and included for reference. The binary was compiled with the \ttfamily -no-pie \normalfont, \ttfamily -fno-pie \normalfont, and \ttfamily -mplt \normalfont compiler flags, causing more frequent use of immediate call instructions.

{\renewcommand{\arraystretch}{1.5}%
\begin{table}
    \centering
    \caption[Binaries used in the analysis]{Binaries used in the analysis.}
    \captionsetup{justification=centering}
        \begin{tabular}{c|c|p{2.6cm}|c|p{3.7cm}}
            \textbf{Program} & \textbf{Architecture} & \textbf{Source} & \textbf{Version} & \textbf{Used for}\\\hline\hline
            cURL & MIPS & \href{https://github.com/darkerego/mips-binaries}{\underline{GitHub}} & Undisclosed & Opcode detection \& OCP-Score evaluation\\\hline
            cURL & Aarch64 & \href{https://curl.se/}{\underline{cURL website}}& 8.0.1 &Opcode detection \& OCP-Score evaluation\\\hline
            cURL & MIPS  & Cross-compiled from \href{https://github.com/curl/curl/releases/tag/curl-8_0_1}{\underline{source}} & 8.0.1& Opcode detection\\\hline
            cURL & x86\_64 & Compiled from \href{https://github.com/curl/curl/releases/tag/curl-8_0_1}{\underline{source}} &8.0.1 & Opcode detection\\\hline
            OpenVPN & MIPS & \href{https://github.com/darkerego/mips-binaries}{\underline{GitHub}} &Undisclosed & Opcode detection \& OCP-Score evaluation\\\hline
            OpenVPN & Aarch64 & \href{https://archlinuxarm.org/packages/aarch64/openvpn}{\underline{Arch repository}}  &2.6.4-1 & Opcode detection \& OCP-Score evaluation\\\hline
            Chipquarium & Chip8 & \href{https://github.com/JohnEarnest/chip8Archive/}{\underline{GitHub}} &1.0 & Call graph \\
    \end{tabular}
    \label{tab:binaries_source}
\end{table}}

The parameters found in Table \ref{tab:parameters_binaries} were obtained by analysing the binaries with command-line tools such as \textbf{readelf}, \textbf{size}, and \textbf{objdump}, and by reading the documentation of the architectures. 

A specific modification was implemented for the MIPS and Aarch64 parameters in this process: the \textbf{pcOffset} and \textbf{pcIncPerInstr} parameters were divided by a value of 4 compared to what their architecture specified for them. This adjustment serves to emulate a left shift operation on the operand of the call instruction by a value of 2, as suggested by the architectural references \cite{aarch64}\cite{mipsisa}.

As mentioned earlier, there is also a cross-compiled binary of cURL for the MIPS architecture, this binary has the same parameters as the cURL MIPS binary, with the exception of \textbf{fileOffsetEnd} which has a value of 567492 instead.

{\renewcommand{\arraystretch}{1.3}%
\begin{table}
    \centering
    \captionsetup{justification=centering}
    \caption[API parameters used in the analysis]{API parameters used in the analysis.}
    \hspace*{-1cm}
        \begin{tabular}{p{3.5cm}||p{1.6cm}|p{1.6cm}|p{1.6cm}|p{1.6cm}|p{1.6cm}|p{1.6cm}}
            \diagbox[width=3.6cm, height=1.4cm]{\textbf{Parameters}}{\textbf{Binaries}} & \textbf{cURL MIPS} & \textbf{cURL Aarch64} & \textbf{cURL x86\_64} & \textbf{\small{OpenVPN} MIPS}  & \textbf{\small{OpenVPN} Aarch64 }& \textbf{Chipquarium Chip8} \\\hline\hline
            instructionLength                   & 32 & 32 & 32 & 32 & 32 & 16\\\hline
            retOpcodeLength                     & 32 & 32 & 8 & 32 & 32 & 16 \\\hline
            callOpcodeLength                    & 6 & 6 & 8 & 6 & 6 & 4 \\\hline
            fileOffset                          & 0 & 4096 & 0 & 0 & 68416 & 0 \\\hline
            fileOffsetEnd                       & 94560 & 2163136 & 501176 & 1782196 & 753456 & 1072 \\\hline
            pcOffset                            & \ttfamily 0x100000 \normalfont & ANY & \ttfamily 0x100000 \normalfont &  \ttfamily 0x100000 \normalfont & ANY & \ttfamily 0x200 \normalfont \\\hline
            pcIncPerInstr                       & 1 & 1 & 1 & 1 & 1 & 2  \\\hline
            endiannes                           & "big" & "little" & "little & "big" & "little" & "big"  \\\hline
            nrCandidates                        & 5 & 5 & 5 & 5 & 5 & 5  \\\hline
            callCandidateRange                  & 0, 20 & 0, 20 & 0, 20 & 0, 20  & 0, 20 & 0, 20  \\\hline
            retCandidateRange                   & 0, 10 & 0, 10 & 0, 10 & 0, 10 & 0, 10 & 0, 10 \\\hline
            returnToFunction-PrologueDistance   & 3 & 3 & 3 & 3 & 3 & 3  \\\hline
            unknownCodeEntry                    & False & False & False & False & False & False  \\\hline
            includeInstructions                 & False & False & False & False & False & False\\\hline
            isRelativeAddressing                & False & True & False & False & True & False\\
        \end{tabular}
    \hspace*{-1cm}
    \label{tab:parameters_binaries}
\end{table}}

\subsection{Return and call opcode detection}
Tables \ref{tab:openvpn_mips}, \ref{tab:openvpn_aarch}, \ref{tab:curl_mips}, and \ref{tab:curl_aarch} present the top five probable candidates for call and return opcodes for the OpenVPN MIPS, OpenVPN Aarch64, cURL MIPS, and cURL Aarch64 binaries, respectively. The correctly identified opcodes emerge as most probable with a substantial margin in Tables \ref{tab:openvpn_mips} and \ref{tab:curl_aarch}, whereas the remaining two tables reveal contrasting outcomes.

Upon examining the binary in Table \ref{tab:openvpn_aarch}, it is observed that the call instruction appears approximately 1600 times. However, roughly 1200 of these instances are deemed invalid as they lack a preceding return instruction above the called function. The NOP instruction (\ttfamily{0xD503201F}\normalfont) frequently precedes function prologues in this binary, which accounts for its higher OCP-Score as a potential return opcode.

The results also differ for the binary featured in Table \ref{tab:curl_mips}. In this case, the call instruction and return instruction are encountered about 40 and 200 times, respectively. The return instruction does not rank within the top 20 instructions, and as a result, it falls outside the predefined search range defined by the \textbf{retCandidateRange} parameter. Despite this, the opcode associated with the branch instruction, \ttfamily{0x08}\normalfont, is assigned a OCP-Score of roughly 0.4.

\def\checkmark{\tikz\fill[scale=0.4](0,.35) -- (.25,0) -- (1,.7) -- (.25,.15) -- cycle;} 
\begin{minipage}{.45\textwidth}
    
{\renewcommand{\arraystretch}{1.5}%
\begin{table}[H]
    \centering
    \caption[Top 5 most probable candidates - OpenVPN MIPS]{Top 5 most probable return and call opcodes from the OpenVPN binary with MIPS architecture.}
    \scriptsize\begin{tabular}{c|c|c|c}
        \textbf{OCP-Score} & \textbf{Call opcode} & \textbf{Return opcode} & \textbf{Correct} \\\hline\hline
        0.866 & \tt{0x0C000000} & \tt{0x03E00008} & \checkmark  \\\hline
        0.449 & \tt{0x08000000} & \tt{0x0320F809} &  \\\hline
        0.412 & \tt{0x08000000} & \tt{0x8FBC0018} &  \\\hline
        0.388 & \tt{0x08000000} & \tt{0xAFA20010} &  \\\hline
        0.373 & \tt{0x08000000} & \tt{0x00001021} &  \\
    \end{tabular}
    \label{tab:openvpn_mips}
\end{table}}

\end{minipage}%
\hfill\begin{minipage}{.45\textwidth}

{\renewcommand{\arraystretch}{1.5}%
\begin{table}[H]
    \centering
    \caption[Top 5 most probable candidates - OpenVPN Aarch64]{Top 5 most probable return and call opcodes from the OpenVPN binary with Aarch64 architecture.}
    \scriptsize\begin{tabular}{c|c|c|c}
        \textbf{OCP-Score} & \textbf{Call opcode} & \textbf{Return opcode} & \textbf{Correct} \\\hline\hline
        0.612 & \tt{0x94000000} & \tt{0xD503201F} &  \\\hline
        0.478 & \tt{0x94000000} & \tt{0xD65F03C0} & \checkmark  \\\hline
        0.426 & \tt{0x94000000} & \tt{0xD63F0060} &  \\\hline
        0.398 & \tt{0x14000000} & \tt{0xD63F0060} &  \\\hline
        0.396 & \tt{0x14000000} & \tt{0x72001C1F} &  \\
    \end{tabular}
    \label{tab:openvpn_aarch}
\end{table}}
    
\end{minipage}

\begin{minipage}{.45\textwidth}
{\renewcommand{\arraystretch}{1.5}%
\begin{table}[H]
    \centering
    \caption[Top 5 most probable candidates - cURL MIPS]{Top 5 most probable return and call opcodes from the cURL binary with MIPS architecture.}
    \scriptsize\begin{tabular}{c|c|c|c}
        \textbf{OCP-Score} & \textbf{Call opcode} & \textbf{Return opcode} & \textbf{Correct} \\\hline\hline
        0.389 & \tt{0x08000000} & \tt{0x8FBC0010} &  \\\hline
        0.376 & \tt{0x08000000} & \tt{0x8FBC0020} &  \\\hline
        0.368 & \tt{0x0C000000} & \tt{0x8FBC0010} &  \\\hline
        0.365 & \tt{0x08000000} & \tt{0x8FBC0018} &  \\\hline
        0.357 & \tt{0x08000000} & \tt{0x0320F809} &  \\
    \end{tabular}
    \label{tab:curl_mips}
\end{table}}
\end{minipage}%
\hfill\begin{minipage}{.45\textwidth}

{\renewcommand{\arraystretch}{1.5}%
\begin{table}[H]
    \centering
    \caption[Top 5 most probable candidates - cURL Aarch64]{Top 5 most probable return and call opcodes from the cURL binary with Aarch64 architecture.}
    \scriptsize\begin{tabular}{c|c|c|c}
        \textbf{OCP-Score} & \textbf{Call opcode} & \textbf{Return opcode} & \textbf{Correct} \\\hline\hline
        0.698 & \tt{0x94000000} & \tt{0xD65F03C0} & \checkmark \\\hline
        0.367 & \tt{0x94000000} & \tt{0xA94153F3} &  \\\hline
        0.353 & \tt{0x14000000} & \tt{0xD65F03C0} &  \\\hline
        0.346 & \tt{0x94000000} & \tt{0x52800020} &  \\\hline
        0.334 & \tt{0x94000000} & \tt{0xAA1303E0} &  \\
    \end{tabular}
    \label{tab:curl_aarch}
\end{table}}
\end{minipage}

As mentioned earlier, an additional binary for cURL MIPS was cross-compiled with additional compiler flags enabled, to ensure an appropriate frequency of immediate call instructions. The results for this binary, along with the x86\_64 binary, which uses a variable-length instruction format, can be seen in Tables \ref{tab:curl_mips_compiled} and \ref{tab:curl_x86}, respectively.

\begin{minipage}{.45\textwidth}

{\renewcommand{\arraystretch}{1.5}%
\begin{table}[H]
    \centering
    \caption[Top 5 most probable candidates - Cross-compiled cURL MIPS]{Top 5 most probable return and call opcodes from the cross-compiled cURL binary with MIPS architecture.}
    \scriptsize\begin{tabular}{c|c|c|c}
        \textbf{OCP-Score} & \textbf{Call opcode} & \textbf{Return opcode} & \textbf{Correct} \\\hline\hline
        0.598 & \tt{0x0C000000} & \tt{0x03E00008} & \checkmark  \\\hline
        0.378 & \tt{0x0C000000} & \tt{0x00001025} &  \\\hline
        0.345 & \tt{0x0C000000} & \tt{0x00002825} &  \\\hline
        0.342 & \tt{0x0C000000} & \tt{0x24020001} &  \\\hline
        0.340 & \tt{0x0C000000} & \tt{0x02002025} &  \\
    \end{tabular}
    \label{tab:curl_mips_compiled}
\end{table}}

\end{minipage}%
\hfill\begin{minipage}{.45\textwidth}
{\renewcommand{\arraystretch}{1.5}%
\begin{table}[H]
    \centering
    \caption[Top 5 most probable candidates - cURL x86\_64]{Top 5 most probable return and call opcodes from the cURL binary with x86\_64 architecture.}
    \scriptsize\begin{tabular}{c|c|c|c}
        \textbf{OCP-Score} & \textbf{Call opcode} & \textbf{Return opcode} & \textbf{Correct} \\\hline\hline
        0.001 & \tt{0xF00000000} & \tt{0x4800000000} &  \\\hline
        0.001 & \tt{0xF00000000} & \tt{0x8B00000000} &  \\\hline
        0.001 & \tt{0xF00000000} & \tt{0xFF00000000} &  \\\hline
        0.001 & \tt{0xF00000000} & \tt{0x2400000000} &  \\\hline
        0.001 & \tt{0xF00000000} & \tt{0x8900000000} &  \\
    \end{tabular}
    \label{tab:curl_x86}
\end{table}}

\end{minipage}

\subsection{OCP-Score as a metric}
\begin{figure}
    \centering
    \begin{subfigure}[b]{0.49\textwidth}
        \centering
        \includegraphics[width=\textwidth]{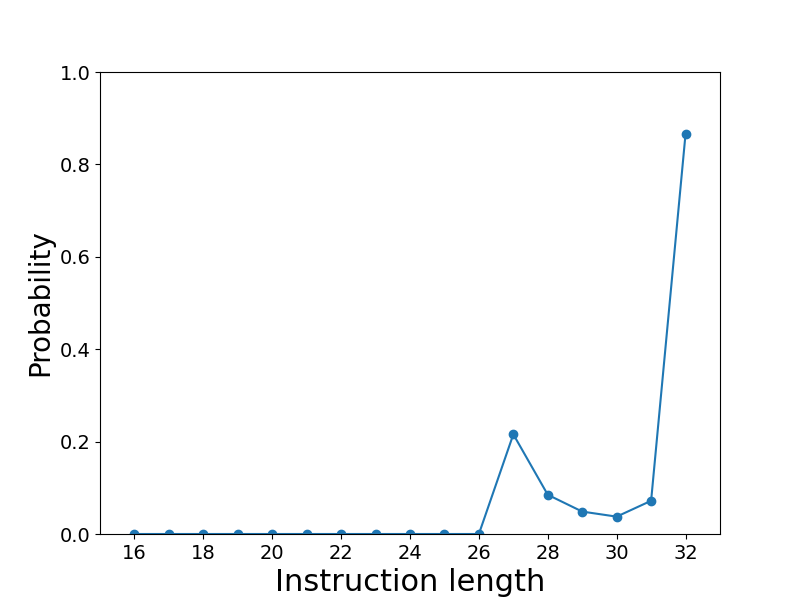}
        \caption{{\small OpenVPN in MIPS architecture}}   
    \end{subfigure}
    \hfill
    \begin{subfigure}[b]{0.49\textwidth}  
        \centering 
        \includegraphics[width=\textwidth]{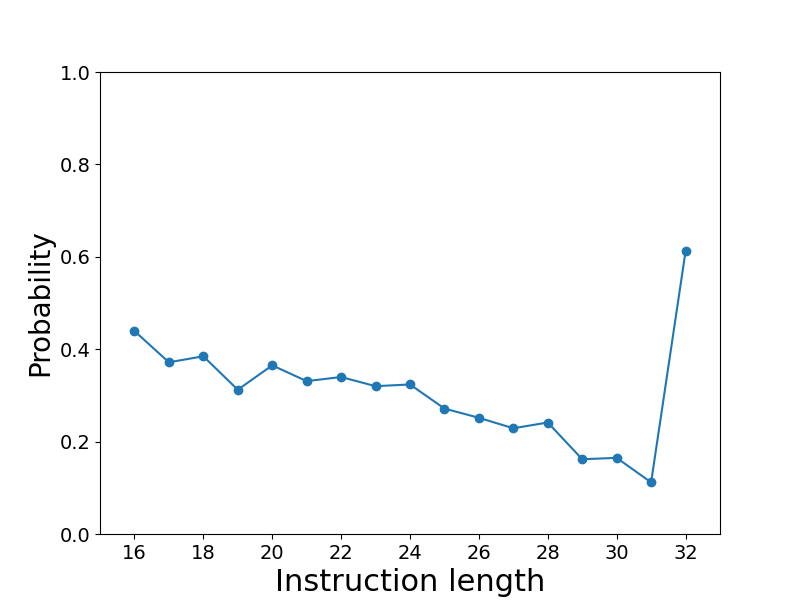}
        \caption[]%
        {{\small OpenVPN in Aarch64 architecture}}    
    \end{subfigure}
    \vskip\baselineskip
    \begin{subfigure}[b]{0.49\textwidth}   
        \centering 
        \includegraphics[width=\textwidth]{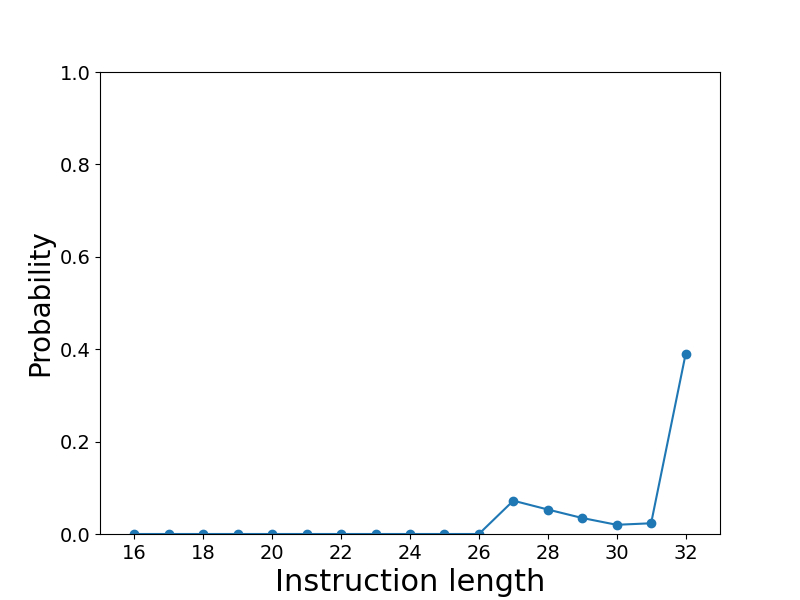}
        \caption[]%
        {{\small cURL in MIPS architecture}}    
    \end{subfigure}
    \hfill
    \begin{subfigure}[b]{0.49\textwidth}   
        \centering 
        \includegraphics[width=\textwidth]{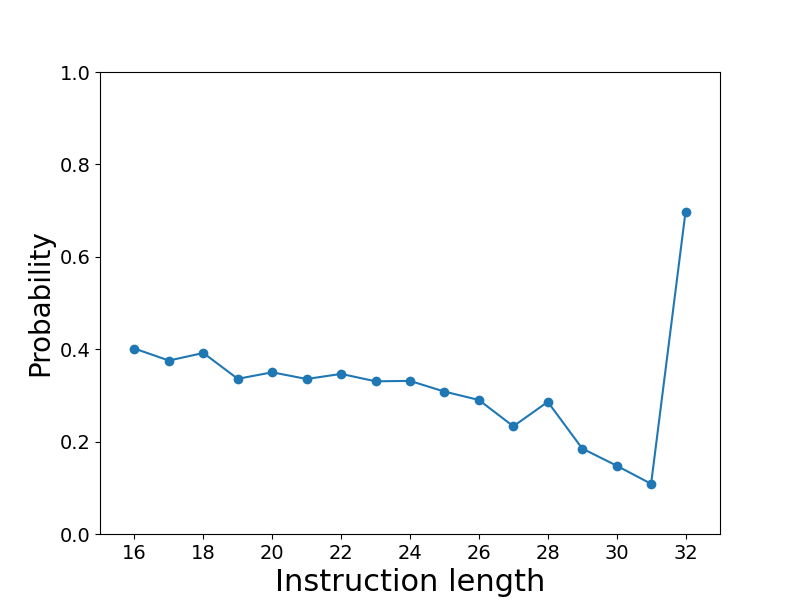}
        \caption[]%
        {{\small cURL in Aarch64 architecture}}    
    \end{subfigure}
    \caption[Instruction length parameter and OCP-Score]{OCP-Score for different inputs of the \textit{instructionLength} parameter, shown for the cURL and OpenVPN binaries in the MIPS and Aarch64 architectures.} 
    \label{fig:plot_instr_prob}
\end{figure}

%%%%%%%%%%%%%%%%%%%%%%%%
% INSTRUCTION LENGTH
%%%%%%%%%%%%%%%%%%%%%%%%
Figure \ref{fig:plot_instr_prob} depicts the maximum OCP-Score corresponding to various values of the instruction length variable. The MIPS binaries exhibit a low OCP-Score for all values except the correct one. In contrast, the Aarch64 architecture binaries display greater variability, with higher OCP-Score for incorrect values.

This discrepancy may arise due to the differing addressing modes employed in the call instructions. In the MIPS architecture with absolute addressing, a valid call operand must be an address in the range between the first and last instruction, for instance, within the range of \ttfamily 0x400160 \normalfont and \ttfamily 0x5B3290 \normalfont in the case of the OpenVPN MIPS binary. Conversely, a relative call instruction may involve lower values, which are arguably more common in noisy data. For example, an operand value of 4 would point toward the instruction preceding the call instruction itself.

\begin{figure}
    \centering
    \begin{subfigure}[b]{0.49\textwidth}
        \centering
        \includegraphics[width=\textwidth]{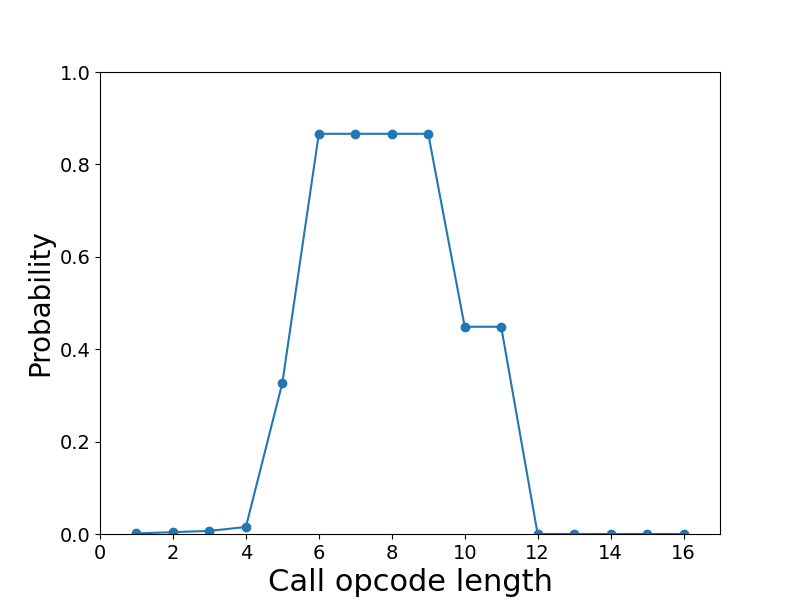}
        \caption{{\small OpenVPN in MIPS architecture}}   
    \end{subfigure}
    \hfill
    \begin{subfigure}[b]{0.49\textwidth}  
        \centering 
        \includegraphics[width=\textwidth]{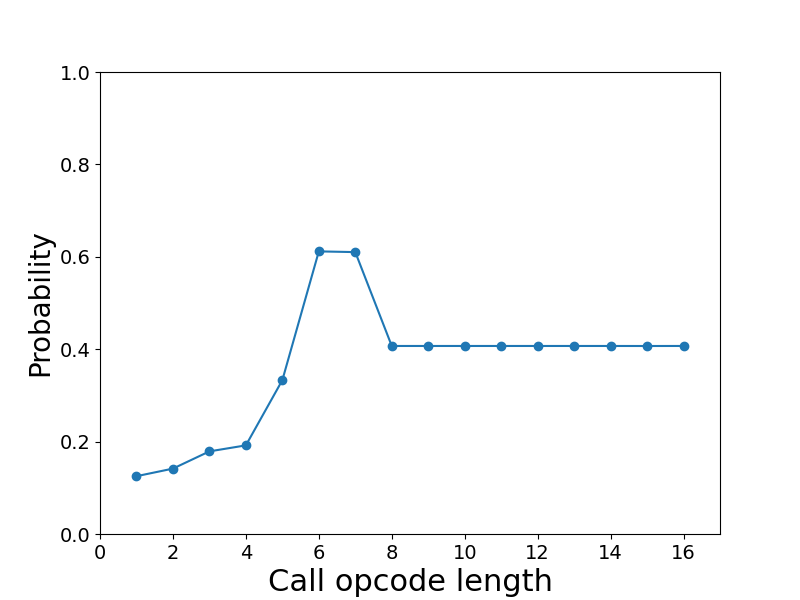}
        \caption[]%
        {{\small OpenVPN in Aarch64 architecture}}    
    \end{subfigure}
    \vskip\baselineskip
    \begin{subfigure}[b]{0.49\textwidth}   
        \centering 
        \includegraphics[width=\textwidth]{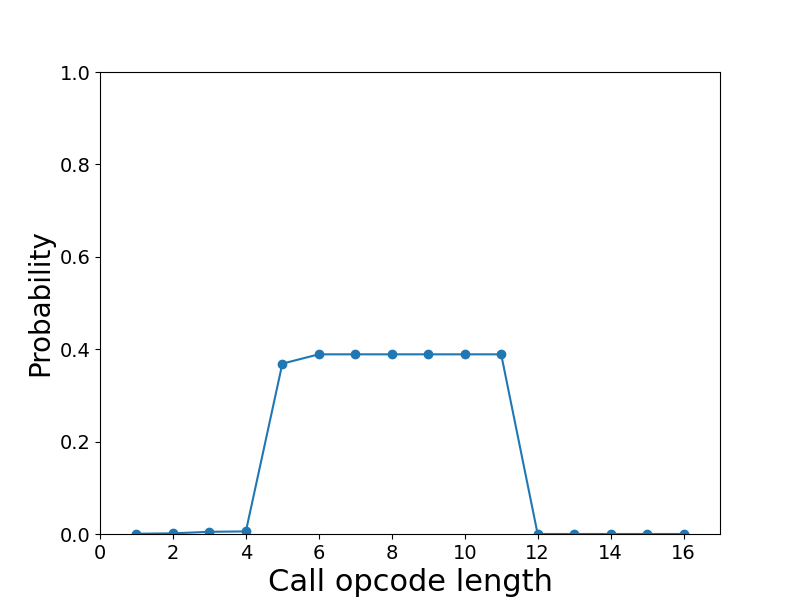}
        \caption[]%
        {{\small cURL in MIPS architecture}}    
    \end{subfigure}
    \hfill
    \begin{subfigure}[b]{0.49\textwidth}   
        \centering 
        \includegraphics[width=\textwidth]{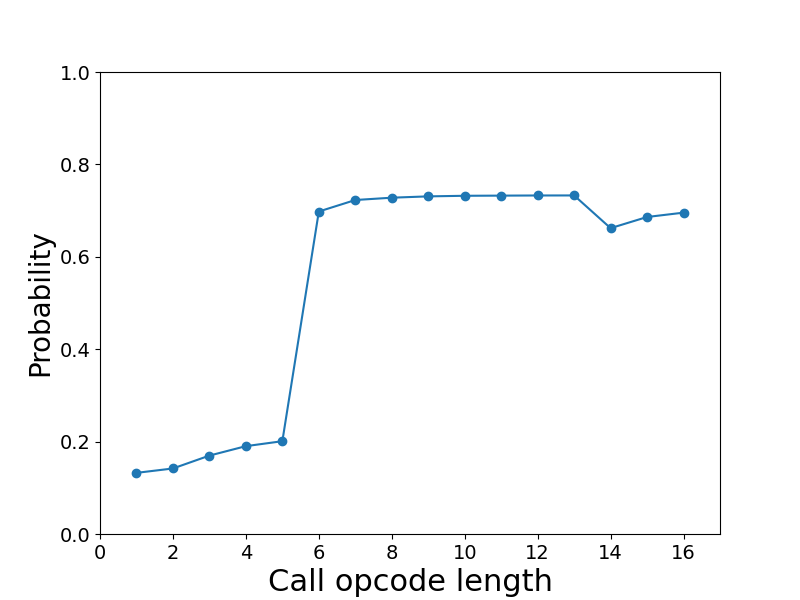}
        \caption[]%
        {{\small cURL in Aarch64 architecture}}    
    \end{subfigure}
    \caption[Call opcode length parameter and OCP-Score]{OCP-Score for different inputs of the \textit{callOpcodeLength} parameter, shown for the cURL and OpenVPN binaries in the MIPS and Aarch64 architectures.} 
    \label{fig:plot_call_prob}
\end{figure}

%%%%%%%%%%%%%%%%%%%%%%%%
% CALL OPCODE LENGTH
%%%%%%%%%%%%%%%%%%%%%%%%
Figure \ref{fig:plot_call_prob} depicts the maximum OCP-Score corresponding to various values of call opcode length. The data suggests that multiple values close to the correct value give a high OCP-Score. The explanation for this is presented in Section \ref{discussion}. 
%This can be explained by the fact that the most significant bits are rarely used, for instance when calling an instruction less than 64 Mb away from the callee, only the 14 least significant bits hold information. Specifically, the rest of the bits are set to 0 for absolute and positive relative calls, and to 1 for negative relative calls, due to the operand being a signed integer.

\begin{figure}
    \centering
    \begin{subfigure}[b]{0.49\textwidth}
        \centering
        \includegraphics[width=\textwidth]{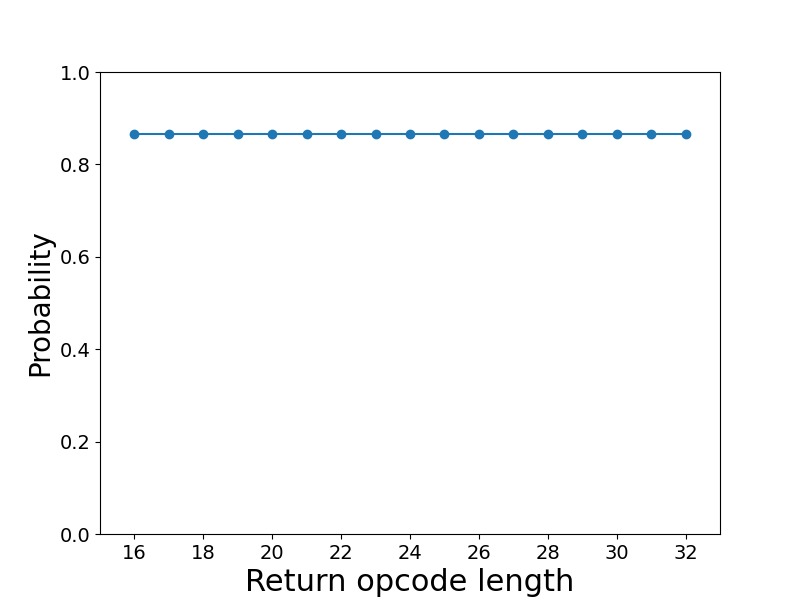}
        \caption{{\small OpenVPN in MIPS architecture}}   
    \end{subfigure}
    \hfill
    \begin{subfigure}[b]{0.49\textwidth}  
        \centering 
        \includegraphics[width=\textwidth]{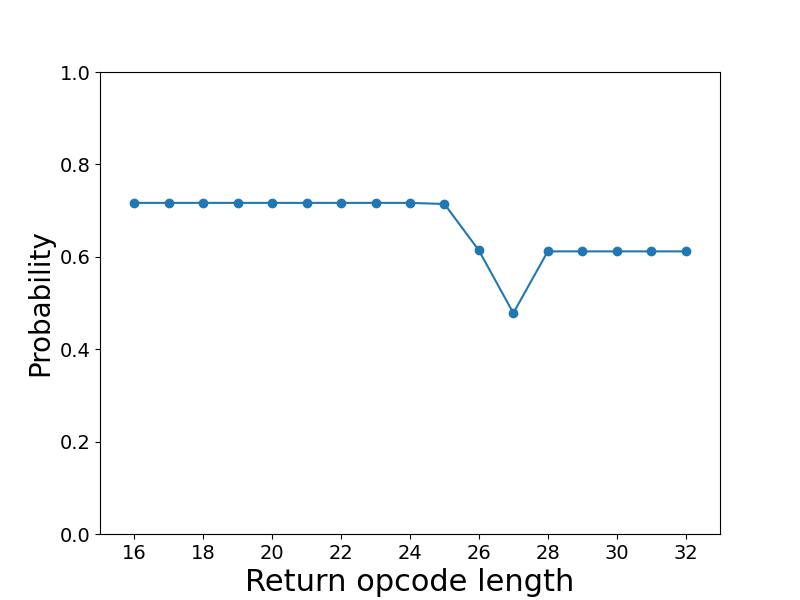}
        \caption[]%
        {{\small OpenVPN in Aarch64 architecture}}    
    \end{subfigure}
    \vskip\baselineskip
    \begin{subfigure}[b]{0.49\textwidth}   
        \centering 
        \includegraphics[width=\textwidth]{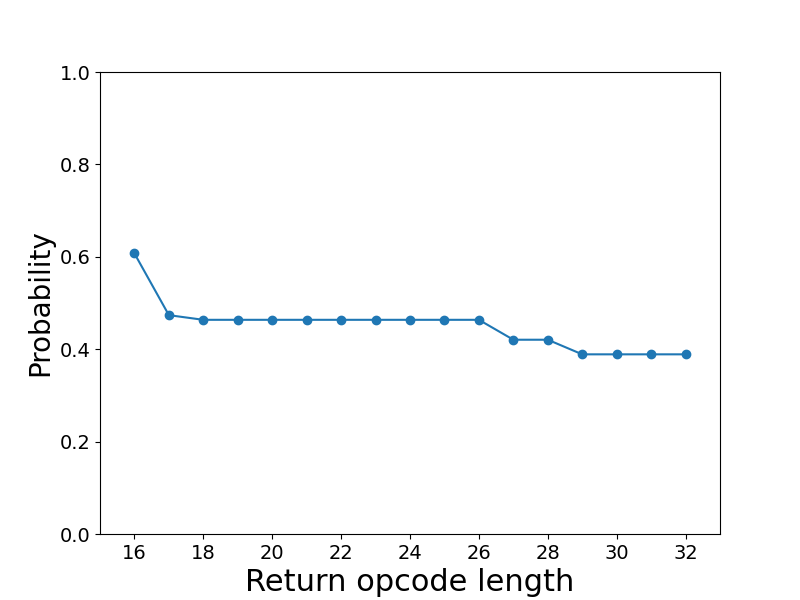}
        \caption[]%
        {{\small cURL in MIPS architecture}}    
    \end{subfigure}
    \hfill
    \begin{subfigure}[b]{0.49\textwidth}   
        \centering 
        \includegraphics[width=\textwidth]{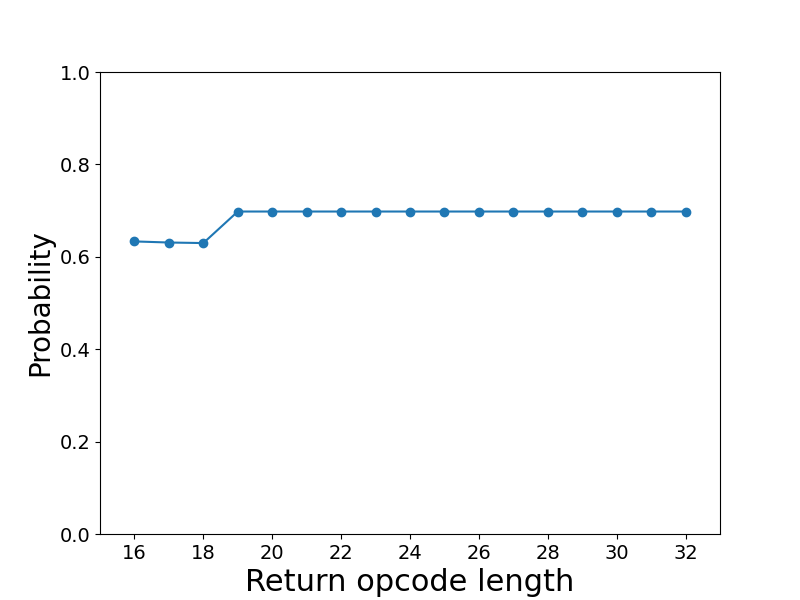}
        \caption[]%
        {{\small cURL in Aarch64 architecture}}    
    \end{subfigure}
    \caption[Return opcode length parameter and OCP-Score]{OCP-Score for different inputs of the \textit{retOpcodeLength} parameter, shown for the cURL and OpenVPN binaries in the MIPS and Aarch64 architectures.} 
    \label{fig:plot_ret_prob}
\end{figure}

%%%%%%%%%%%%%%%%%%%%%%%%
% RETURN OPCODE LENGTH
%%%%%%%%%%%%%%%%%%%%%%%%
Figure \ref{fig:plot_ret_prob} depicts the maximum OCP-Score corresponding to various values of return opcode length. Looking at the data it seems that the change in value is not notably significant between different values. In general, when decreasing the return opcode length, we either see an increase in OCP-Score due to the set of instructions considered to be a return instruction increasing, or a decrease due to another incorrect but more frequent set pushing it out of the return search range.

\begin{figure}
    \centering
    \begin{subfigure}[b]{0.49\textwidth}
        \centering
        \includegraphics[width=\textwidth]{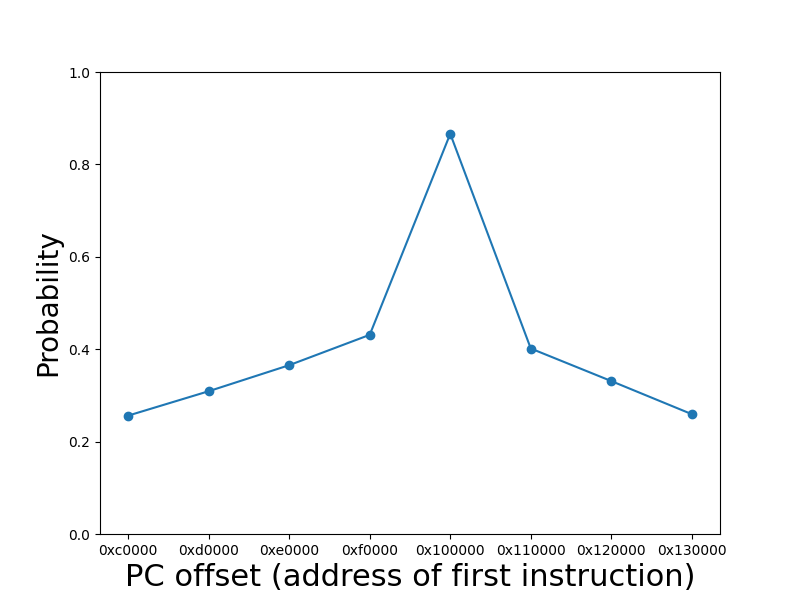}
        \caption{{\small OpenVPN in MIPS architecture}}   
    \end{subfigure}
    \hfill
    \begin{subfigure}[b]{0.49\textwidth}  
        \centering 
        \includegraphics[width=\textwidth]{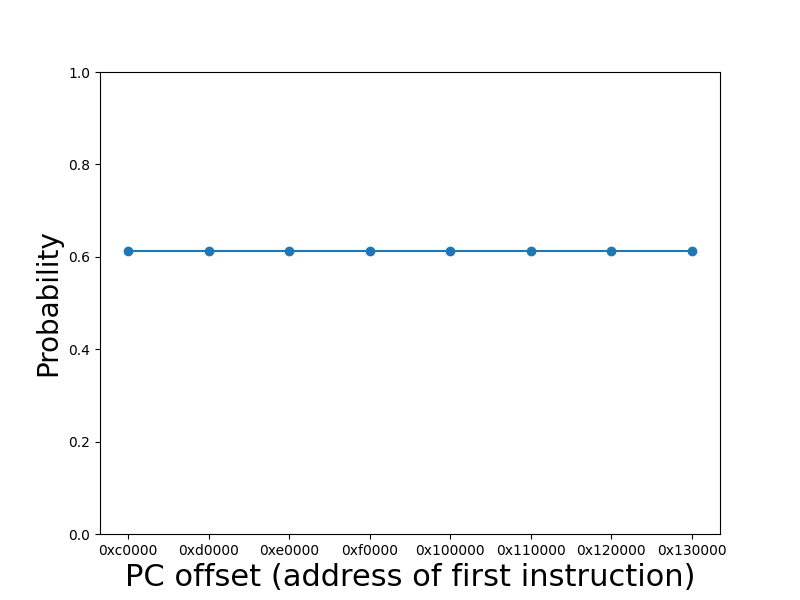}
        \caption[]%
        {{\small OpenVPN in Aarch64 architecture}}    
    \end{subfigure}
    \vskip\baselineskip
    \begin{subfigure}[b]{0.49\textwidth}   
        \centering 
        \includegraphics[width=\textwidth]{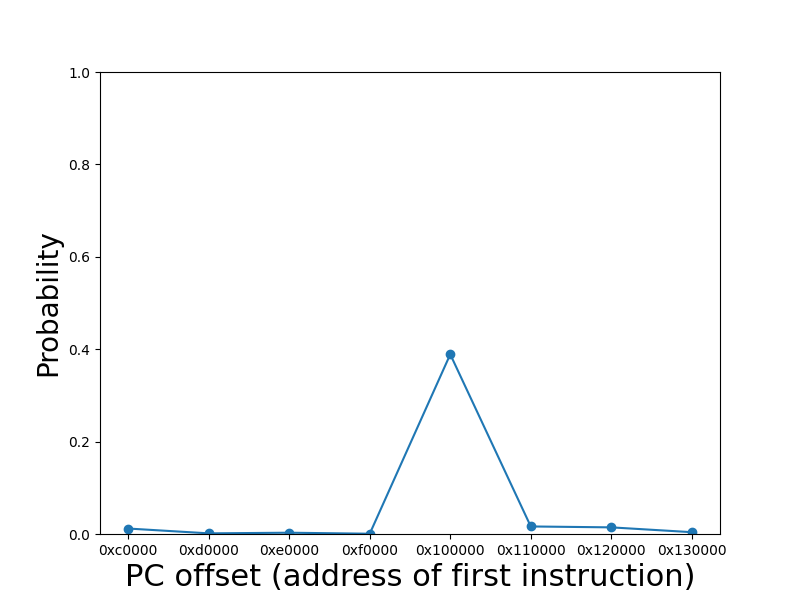}
        \caption[]%
        {{\small cURL in MIPS architecture}}    
    \end{subfigure}
    \hfill
    \begin{subfigure}[b]{0.49\textwidth}   
        \centering 
        \includegraphics[width=\textwidth]{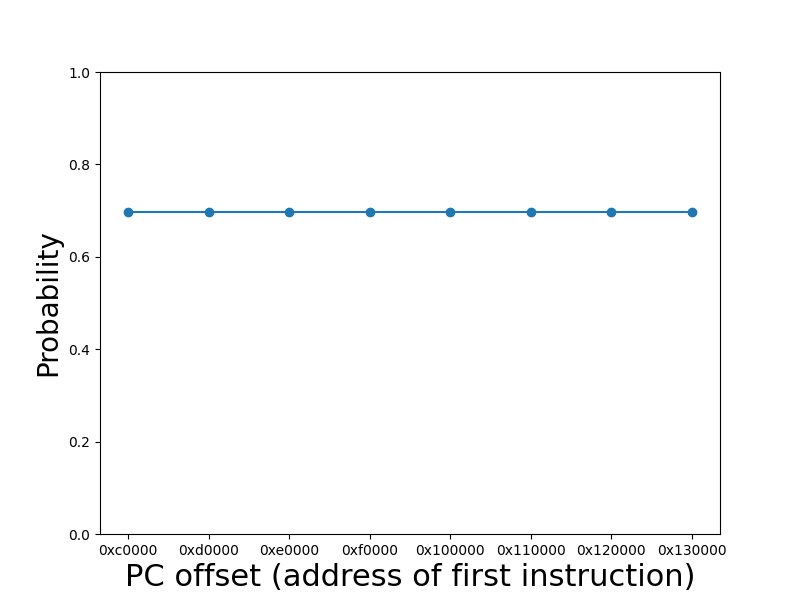}
        \caption[]%
        {{\small cURL in Aarch64 architecture}}    
    \end{subfigure}
    \caption[PC offset parameter and OCP-Score ]{OCP-Score for different inputs of the \textit{pcOffset} parameter, shown for the cURL and OpenVPN binaries in the MIPS and Aarch64 architectures.}
    \label{fig:plot_pcoffset_prob}
\end{figure}

%%%%%%%%%%%%%%%%%%%%%%%%
% PC OFFSET
%%%%%%%%%%%%%%%%%%%%%%%%
Figure \ref{fig:plot_pcoffset_prob} depicts the maximum OCP-Score corresponding to various values of PC offset. It is important to clarify that these values do not affect the particular instructions read from the binary file, but rather assign a specific address to each instruction. For example, with a PC offset value of \ttfamily 0x1000\normalfont, the first instruction would be given an address of \ttfamily 0x1000\normalfont. From the results, it is evident that the PC offset value has no impact on relative addressing, which aligns with expectations. However, in the context of absolute addressing in MIPS, the correct value gives a significantly higher OCP-Score (subfigures (a) and (c)).

\begin{figure}
    \centering
    \begin{subfigure}[b]{0.49\textwidth}
        \centering
        \includegraphics[width=\textwidth]{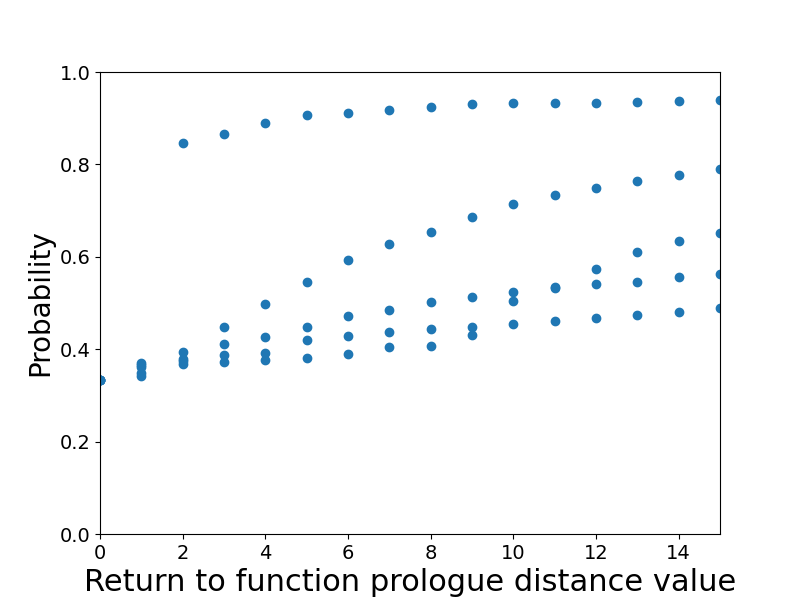}
        \caption{{\small OpenVPN in MIPS architecture}}   
    \end{subfigure}
    \hfill
    \begin{subfigure}[b]{0.49\textwidth}  
        \centering 
        \includegraphics[width=\textwidth]{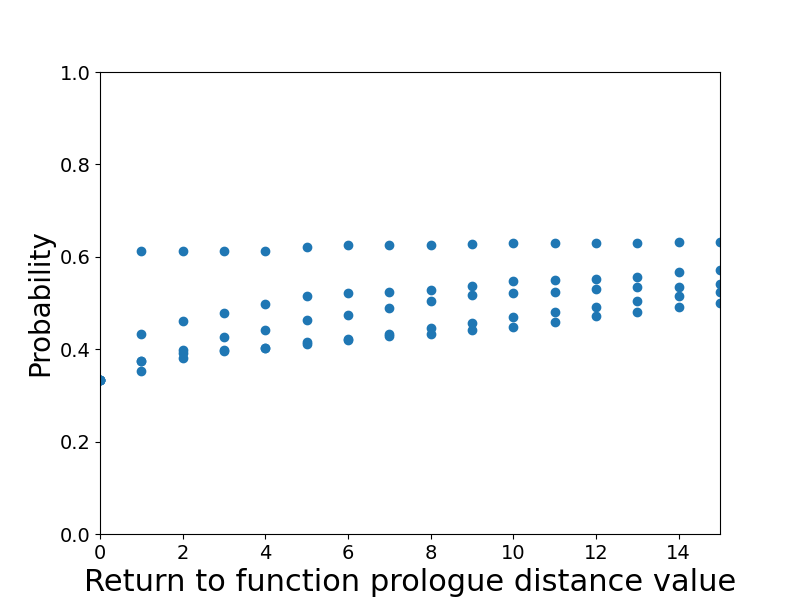}
        \caption[]%
        {{\small OpenVPN in Aarch64 architecture}}    
    \end{subfigure}
    \vskip\baselineskip
    \begin{subfigure}[b]{0.49\textwidth}   
        \centering 
        \includegraphics[width=\textwidth]{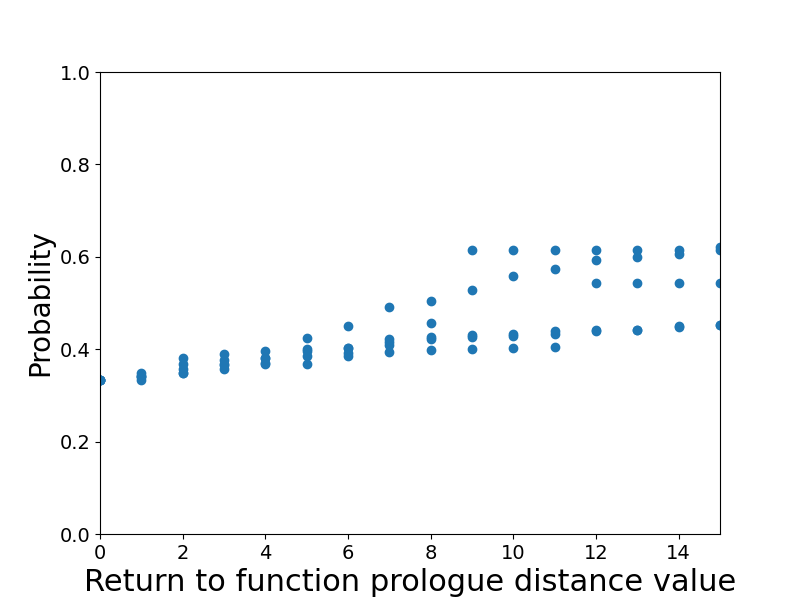}
        \caption[]%
        {{\small cURL in MIPS architecture}}    
    \end{subfigure}
    \hfill
    \begin{subfigure}[b]{0.49\textwidth}   
        \centering 
        \includegraphics[width=\textwidth]{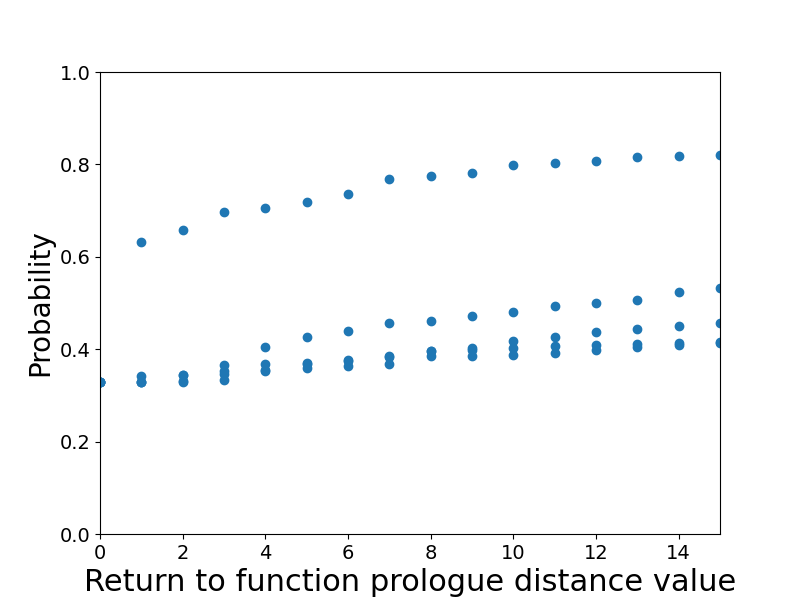}
        \caption[]%
        {{\small cURL in Aarch64 architecture}}    
    \end{subfigure}
    \caption[Return to function distance parameter and OCP-Score]{OCP-Score for different inputs of the \textit{returnToFunctionPrologueDistance} parameter, shown for the cURL and OpenVPN binaries in the MIPS and Aarch64 architectures.}
    \label{fig:plot_retdist_prob}
\end{figure}

%%%%%%%%%%%%%%%%%%%%%%%%
% RETURN TO FUNCTION PROLOGUE DISTANCE
%%%%%%%%%%%%%%%%%%%%%%%%
Figure \ref{fig:plot_retdist_prob} depicts the five highest OCP-Scores corresponding to various values of return to function prologue distance. This value determines how far above a function prologue one can search for a potential return instruction. From the data, we can see that a value of 2 is necessary to correctly detect functions in MIPS, and a value of 1 is sufficient in Aarch64. Values higher than this introduce additional noise in the data, by amplifying the OCP-Score of incorrect opcodes.

\subsection{Call graph creation}
\begin{figure}
    \centering
    \includegraphics[width=0.8\linewidth]{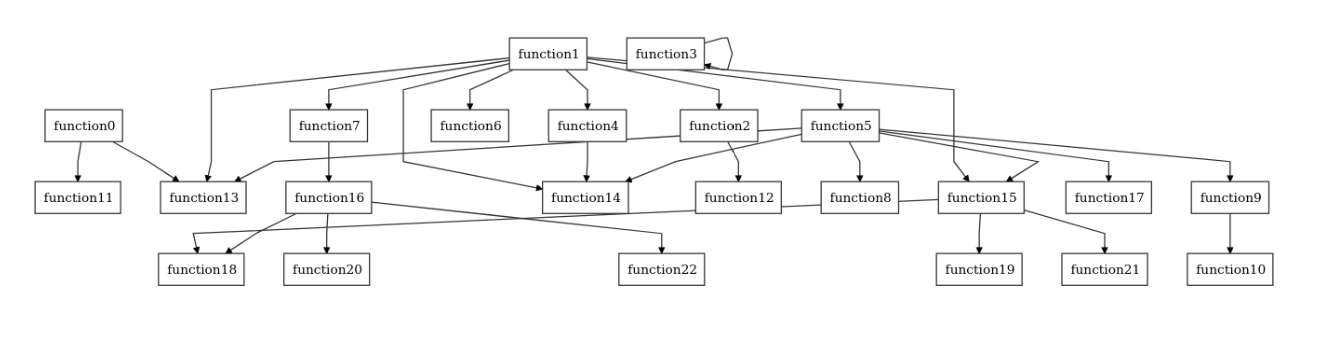}
    \caption[Call graph - source code]{Call graph of the Chipquarium binary hand-crafted from the source code.}
    \label{fig:call_graph_original}
\end{figure}

\begin{figure}
    \centering
    \includegraphics[width=0.8\linewidth]{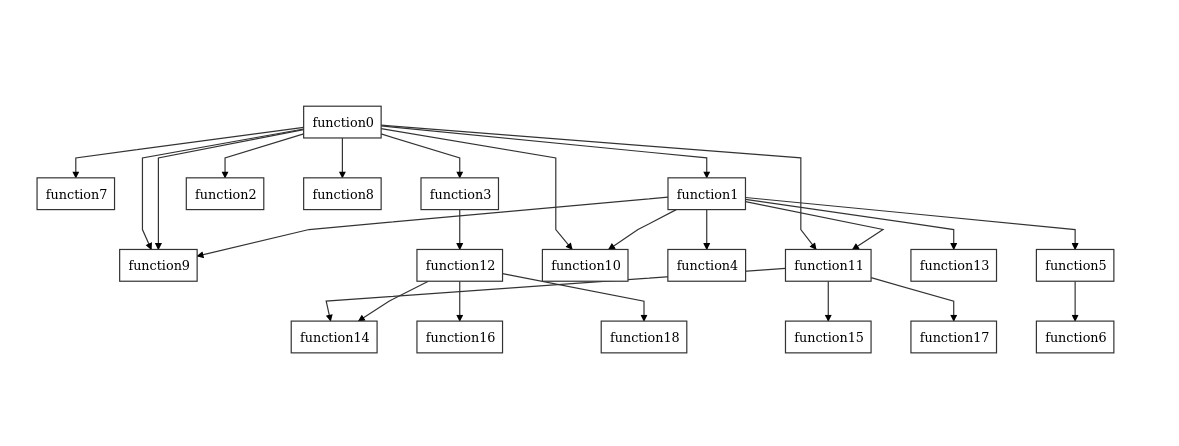}
    \caption[Call graph - source code with merged functions]{Call graph of the Chipquarium binary hand-crafted from the source code with the first five functions merged into one.}
    \label{fig:call_graph_edited}
\end{figure}

\begin{figure}
    \centering
    \includegraphics[width=0.8\linewidth]{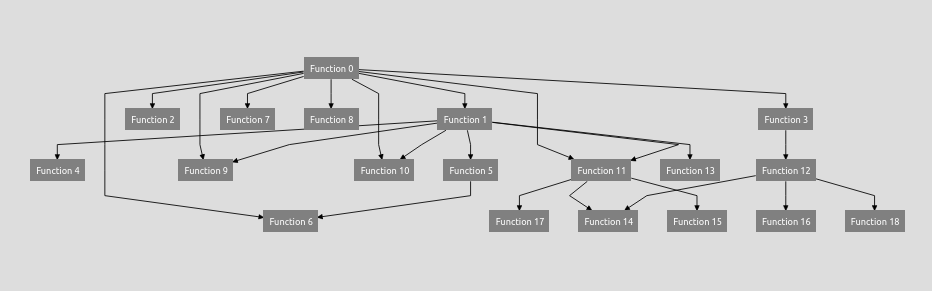}
    \caption[Call graph - generated by our approach]{Call graph of the Chipquarium binary as generated by our approach.}
    \label{fig:call_graph_program}
\end{figure}

To illustrate the call graph functionality effectively, a small program is optimal as it allows clear visualization of the distinct function nodes and edges. In the ensuing figures, different versions of a call graph from the Chipquarium program are presented. Figure \ref{fig:call_graph_original} depicts the call graph derived from inspecting the functions and function calls in the source code. Figure \ref{fig:call_graph_edited} represents the same graph, with the first five functions merged into one, and Figure \ref{fig:call_graph_program} presents the call graph as generated by the developed program. Both Figure \ref{fig:call_graph_edited} and \ref{fig:call_graph_program} showcase identical call graphs, albeit rendered via different graph engines. The rationale behind the merging is due to undetected functions, and will be discussed further in Section \ref{discussion}. 

\section{Discussion}
\label{discussion}
The results presented in the previous section illustrate the application of the OCP-Score, the accuracy of the opcode detection, and the correctness of the created call graph. In this section, we will discuss these results and how they address the research questions.

Starting with the analysis of opcode detection, we can observe that given the binary file has certain properties, such as fixed length instruction format and a significant quantity of return and absolute or relative call instructions, one can effectively distinguish the return and call opcodes from the rest of the instructions. Conversely, a lack of absolute or relative call instructions or a non-fixed length instruction format causes the result to be inconclusive. Therefore, in response to RQ1: it is feasible -- given certain properties and parameters -- to identify the correct call and return instruction. If the results from the analysed binary are inconclusive, this may also provide valuable insight to the reverse engineer: either the provided parameters are incorrect, or the properties of the binary are not what is expected, which can guide subsequent analysis.

An interesting observation from the Aarch64 OpenVPN binary in Table \ref{tab:openvpn_aarch}, was the low frequency of return instruction. However, the program contained a disproportionately high amount of NOP instructions, often found in function epilogues. These instructions have the unique property that they often occur successively, usually more than 3 times. This pattern should make them detectable, and a further improvement to the heuristic approach described in this paper could be to discard them as candidates for call and return instructions, reducing noise in the resulting ranking.

In order to address RQ2, an analysis of the OCP-Score was conducted to determine the effectiveness and limitations of the approach. This analysis iterated over a selection of parameters examining its sensitivity to noise and its impact on the output. Figure \ref{fig:plot_instr_prob} presented how the highest OCP-Score differed with different values for the instruction length parameter. This parameter is unique in that changing its values changes how instructions are extracted, and each value gives a unique output. All values but the correct one generates a list of instructions that is essentially a pseudo-random combination of bits. Out of the 68 total iterations, the OCP-Score was dominant in the four cases where the correct value was chosen for the parameter. This result strengthens the viability and usability of the OCP-Score, indicating that it remains robust against random data.

When iterating over different call opcode lengths, we observe that multiple values resulted in a high OCP-Score. This can be attributed to the fact that the most significant bits of the operand rarely hold information. For instance, for absolute calls and positive relative calls, the most significant bits are usually 0, while for negative relative calls, the value is 1, due to it being a signed integer. An interesting consequence of this is that increasing the call opcode length to a value such as 8 would split the positive and negative relative call instructions into two distinct opcodes, where one of them could have a higher OCP-Score than the correct call opcode with a length of 6. This is where the use of the approach combined with manual inspection would prove useful. An experienced reverse engineer could inspect the instructions and deduce that the value of the operand is a signed integer, and identify the correct call opcode length.

Other parameters such as \textbf{returnToFunctionPrologueDistance} seen in Figure \ref{fig:plot_retdist_prob}, \textbf{callCandidateRange} and \textbf{retCandidateRange} require a minimal value to correctly identify the call and return opcodes, but increasing it further would only increase the noise in the resulting output. As an example of this, setting the \textbf{returnToFunctionPrologueDistance} parameter to a high value would give the branch instruction an OCP-Score similar to the call instruction, since the likelihood of there being a return instruction in any of the in-scope instructions preceding the branch target is very high. Increasing the range of the other two parameters also increases the likelihood of noise in the data, due to increased search space. 

The rationale for developing the OCP-Score was twofold: 1) to present an intuitive ranking of the most probable candidates, and 2) to have a simple scalar value that can be quickly referenced by the reverse engineer. Nonetheless, it is important to be aware of the limitations of the value, and use it in conjunction with a manual inspection of the binary, the outputted call graph, and other analyses, for a better and more complete understanding. For instance, an arbitrary instruction that only occurs a few times, where the presumed operand would target an instruction with a return statement preceding it, would output a very high OCP-Score. However, an experienced user would notice that due to the infrequency of the instruction, it is either not likely to be a call instruction, or at the very least the lack of data points renders it inconclusive.

The final analysis examined the call graphs generated from the Chipquarium binary. The analysis revealed that the generated call graph was identical to the hand-crafted call graph, provided that the first 5 functions were merged into a single function. This illustrates the main limitation our approach has with generating call graphs: if a function never gets called, the heuristic apparoch will not identify it as a function. There are potential ways to remedy this, as most architectures have a distinct function prologue, often involving stack operations. Assuming the approach has accurately identified most of the function prologues, the remaining functions could potentially be identified using techniques such as machine learning or pattern matching.

The results and analysis over multiple architectures and binaries demonstrates the effectiveness of the presented heuristic approach. We are confident that it can serve as a useful tool to help reverse engineers in the process of analysing binary programs from unknown instruction set architectures, and fills a much-needed gap in the current research. Despite the effectiveness of the heuristic approach, it is important to be aware of its limitations and to use it in combination with manual inspection and other techniques, for the best overall results.

\section{Conclusion}
\label{conclusion}
The primary objective of this research was aimed at reducing the effort of reverse engineering binaries from unknown instruction set architectures. The results and discussion focused on evaluating key properties of the heuristic approach including opcode detection and the OCP-Score.

The approach is effective when the binary files align with particular properties such as a fixed-length instruction format and the presence of return and absolute or relative call instructions. The accuracy of opcode detection and the robustness of the OCP-Score in dealing with noisy data were notable outcomes of this study.z

However, several limitations were also found and discussed, most notably that variable-length instructions are not supported as seen with the x86\_64 architecture. Furthermore, it was discussed that an integrated approach, incorporating both automatic processing and manual inspection, is both beneficial and necessary for an optimal result.

Regarding future work, several areas have been identified. First, support for variable-length instructions would enable the method to support a wider variety of unknown ISAs. Next, functionality to detect specific instructions such as NOPs could further reduce noise. Additionally, it was found that branch instructions were often detected as the second most probable call opcode, and a potential improvement would be to detect and include information on such branch instructions. Identification of uncalled functions via prologues/epilogues matching would improve binary code coverage. Lastly, evaluation on a larger dataset of binary programs would help generate a clearer picture of performance over a wider variety of ISAs. %Finally, a graph similarity measure would enable automatic comparison of ground-truth call graphs with call graphs derived by our heuristic approach.

%from a usability perspective, providing sane defaults and optional parameters for the API, thus requiring only a subset of the parameter set, would greatly improve the user experience of the reverse engineer.

%In conclusion, a heuristic approach to extracting call graph information of binaries from unknown instruction set architectures was introduced. While there is room for improvement, the framework addresses a significant gap in the toolset available for such tasks. As research in this field continues, it is expected that more tools focusing on unknown instruction set architectures will be developed.

%\section{References}
\bibliographystyle{plain}
\bibliography{references}

\begin{thebibliography}{10}

\bibitem{aarch64}
Arm a-profile a64 instruction set architecture.
\newblock
  \url{https://developer.arm.com/documentation/ddi0602/2023-03/Base-Instructions/BL--Branch-with-Link-?lang=en}.

\bibitem{mipsisa}
Mips reference sheet.
\newblock
  \url{https://uweb.engr.arizona.edu/~ece369/Resources/spim/MIPSReference.pdf}.

\bibitem{bao2014application}
Chongxi Bao, Domenic Forte, and Ankur Srivastava.
\newblock On application of one-class svm to reverse engineering-based hardware
  trojan detection.
\newblock In {\em Fifteenth International Symposium on Quality Electronic
  Design}, pages 47--54. IEEE, 2014.

\bibitem{2012chernov}
Alexander Chernov and Katerina Troshina.
\newblock Reverse engineering of binary programs for custom virtual machines.
\newblock In {\em ReCon 2012}, 2012.

\bibitem{clemens2015automatic}
John Clemens.
\newblock Automatic classification of object code using machine learning.
\newblock {\em Digital Investigation}, 14:S156--S162, 2015.

\bibitem{wikielf}
Wikimedia Commons.
\newblock Executable and linkable format.
\newblock \url{https://en.wikipedia.org/wiki/Executable_and_Linkable_Format}.
\newblock File: \ttfamily{ELF-layout--en.svg}\normalfont.

\bibitem{hardware}
Marc Fyrbiak, Sebastian Strauss, Christian Kison, Sebastian Wallat, Malte
  Elson, Nikol Rummel, and Christof Paar.
\newblock Hardware reverse engineering: Overview and open challenges.
\newblock {\em 2017 IEEE 2nd International Verification and Security Workshop
  (IVSW)}, 2017.

\bibitem{kairajarvi2020isadetect}
Sami Kairaj{\"a}rvi, Andrei Costin, and Timo H{\"a}m{\"a}l{\"a}inen.
\newblock Isadetect: Usable automated detection of cpu architecture and
  endianness for executable binary files and object code.
\newblock In {\em Proceedings of the Tenth ACM Conference on Data and
  Application Security and Privacy}, pages 376--380, 2020.

\bibitem{kinder2012towards}
Johannes Kinder.
\newblock Towards static analysis of virtualization-obfuscated binaries.
\newblock In {\em 2012 19th Working Conference on Reverse Engineering}, pages
  61--70. IEEE, 2012.

\bibitem{qiu2015identifying}
Jing Qiu, Xiaohong Su, and Peijun Ma.
\newblock Identifying functions in binary code with reverse extended control
  flow graphs.
\newblock {\em Journal of Software: Evolution and Process}, 27(10):793--820,
  2015.

\bibitem{sharif2009automatic}
Monirul Sharif, Andrea Lanzi, Jonathon Giffin, and Wenke Lee.
\newblock Automatic reverse engineering of malware emulators.
\newblock In {\em 2009 30th IEEE Symposium on Security and Privacy}, pages
  94--109. IEEE, 2009.

\bibitem{mipsinstrfig}
Kirat~Pal Singh and Shivani Parmar.
\newblock Design of high performance {MIPS} cryptography processor based on
  {T-DES} algorithm.
\newblock {\em CoRR}, abs/1503.03166, 2015.
\newblock File: \ttfamily{MIPS-instruction-Type.png}\normalfont.

\bibitem{observational}
Daniel Votipka, Seth Rabin, Kristopher Micinski, Jeffrey~S. Foster, and
  Michelle~L. Mazurek.
\newblock An observational investigation of reverse engineers’ process and
  mental models.
\newblock {\em Extended Abstracts of the 2019 CHI Conference on Human Factors
  in Computing Systems}, 2019.

\bibitem{xu2018vmhunt}
Dongpeng Xu, Jiang Ming, Yu~Fu, and Dinghao Wu.
\newblock {VMHunt: A verifiable approach to partially-virtualized binary code
  simplification}.
\newblock In {\em Proceedings of the 2018 ACM SIGSAC Conference on Computer and
  Communications Security}, pages 442--458, 2018.

\end{thebibliography}

\end{document}